\documentclass[AER]{AEA}
\usepackage{graphicx}
\usepackage[letterpaper]{geometry}
\usepackage[english]{babel}
\usepackage[style=chicago-authordate,maxbibnames=10]{biblatex}
\bibliography{knowledge}

\DeclareNameFormat{family-given-bold/given-family-bold/last}{\mkbibbold{%
  \ifboolexpr{test {\ifnumless{\value{listcount}}{2}} or test {\ifnumequal{\value{listcount}}{1}}}
    {\ifgiveninits
       {\usebibmacro{name:family-given}
         {\namepartfamily}
         {\namepartgiveni}
         {\namepartprefix}
         {\namepartsuffix}}
       {\usebibmacro{name:family-given}
         {\namepartfamily}
         {\namepartgiven}
         {\namepartprefix}
         {\namepartsuffix}}%
     \ifboolexpe{%
       test {\ifdefvoid\namepartgiven}
       and
       test {\ifdefvoid\namepartprefix}}
       {}
       {\usebibmacro{name:revsdelim}}}
    {\ifgiveninits
       {\usebibmacro{name:given-family}
         {\namepartfamily}
         {\namepartgiveni}
         {\namepartprefix}
         {\namepartsuffix}}
       {\usebibmacro{name:given-family}
         {\namepartfamily}
         {\namepartgiven}
         {\namepartprefix}
         {\namepartsuffix}}}%
  \usebibmacro{name:andothers}}}

  \DeclareNameAlias{sortname}{family-given-bold/given-family-bold/last}

\usepackage{libertine}
\usepackage{newtxmath}

\usepackage{ctable}
\usepackage[font=sc,labelfont=sc, labelsep=none]{caption}
\usepackage{booktabs}

\newcommand{\medcup}{\mathbin{\scalebox{1.5}{\ensuremath{\cup}}}}%
\usepackage{bm}

\usepackage{url}

\draftSpacing{1.5}

\let\oldabstract\abstract
\let\oldendabstract\endabstract
\makeatletter
\renewenvironment{abstract}
{%
               {\list{}{\addtolength{\leftmargin}{1em} 
                        \listparindent 1.5em%
                        \itemindent    \listparindent%
                        \rightmargin   0em
                        \parsep        \z@ \@plus\p@}%
                \item\relax}%
               {\endlist}%
\oldabstract}
{\oldendabstract}
\makeatother

\begin{document}
\urlstyle{same} 

\title{Creation of Knowledge through Exchanges of Knowledge: Evidence from Japanese Patent Data}
\shortTitle{}
\author{Tomoya Mori and Shosei Sakaguchi
\thanks{Mori: Corresponding author. Institute of Economic Research, Kyoto University, Yoshida-Honmachi, Sakyo-Ku, Kyoto 606-8501, Japan, mori@kier.kyoto-u.ac.jp; Research Institute of Economy, Trade and Industry, 11th floor, Annex, Ministry of Economy, Trade and Industry, 1-3-1 Kasumigaseki Chiyoda-Ku, Tokyo 100-8901, Japan.
Sakaguchi: Department of Economics, University College London, Gower Street, London WC1E 6BT, UK, s.sakaguchi@ucl.ac.uk. Acknowledgements: Earlier versions of this paper were presented at Jinan University, Keio University, Kyoto University, RIETI, the University of Tokyo, the 32nd ARSC meeting at Nansan University, the 13th UEA meeting at Columbia University, the 2nd CURE meeting at Singapore Management University, and the Workshop on Innovation, Technological Change, and International Trade at the Technical University of Munich. The authors are particularly indebted to Marcus Berliant, Jorge De la Roca, Gilles Duranton, Masahisa Fujita, Tadao Hoshino, Wen-Tai Hsu, Daishoku Kanehara, Shin Kanaya, Kentaro Nakajima, Ryo Nakajima, Yoshihiko Nishiyama, Ryo Okui, Akihisa Shibata, and Jens Wrona for their constructive comments. This research was conducted as part of the project ``An empirical framework for studying spatial patterns and causal relationships of economic agglomeration'' undertaken at RIETI. It received partial financial support from JSPS Grants-in-Aid for Scientific Research grant numbers 15H03344, 16K13360, and 17H00987.}}
\date{\today}
\pubMonth{}
\pubYear{}
\pubVolume{}
\pubIssue{}
\JEL{D83, D85, O31, R11, C33, C36}
\Keywords{Knowledge creation, Collaboration, Differentiated knowledge, Patents, Technological novelty, Network}

\begin{abstract}
\noindent This study shows evidence for collaborative knowledge creation among individual researchers through direct exchanges of their mutual differentiated knowledge.
Using patent application data from Japan, the collaborative output is evaluated according to the quality and novelty of the developed patents, which are measured in terms of forward citations and the order of application within their primary technological category, respectively.
Knowledge exchange is shown to raise collaborative productivity more through the extensive margin (i.e., the number of patents developed) in the quality dimension, whereas it does so more through the intensive margin in the novelty dimension (i.e., novelty of each patent).
\end{abstract}

\maketitle

\noindent Knowledge is a key element in various aspects of economic modeling. The theoretical development of innovation-driven economic growth dates back at least to \textcite{Shell-AER1966, Shell-Book1967} and \textcite{Romer-JPE1990,Grossman-Helpman-Book1991,Aghion-Howitt-ECMA1992}.
Formalization has focused on different aspects of knowledge, such as the tension between learning-by-doing and innovation \parencite[e.g.,][]{Klette-Kortum-JPE2004}, spillovers \parencite[e.g.,][]{Jovanovic-Rob-REStud1989}, and imitations 
\parencite[e.g.,][]{Perla-Tonetti-JPE2014}.
Some have described the mechanism of knowledge creation.
In \textcite{Weitzman-QJE1998}, extant ideas induce the development of new ideas if recombined with other existing ideas. 
\textcite{Olsson-JEG2000,Olsson-JEG2005} formalized the recombination of ideas by their convex combinations.
\textcite{Akcigit-et-al-NBER2018} considered team formation by endogenous specialization between a leader and members of a team from ex ante symmetric agents.

This study focuses on the theory of \textcite{Berliant-Fujita-IER2008} with regards to the mechanism of collaborative knowledge creation. 
What facilitates collaboration is the exchange of mutual differentiated knowledge between collaborators through their common knowledge.
The key mechanics underlying their theory is that the relative size of common and differentiated knowledge varies depending on the duration of collaboration. 
A longer duration of collaboration increases common knowledge whereas it decreases differentiated knowledge between collaborators, while the opposite is true between non-collaborators.
To maintain balance between the two types of knowledge with their collaborators, inventors seek polyadic collaborations and rotate their collaborators.

Extant empirical studies have primarily focused on knowledge spillovers associated with R\&D expenditure or transactions by firms and industries \parencite[e.g.,][]{Griliches-BJ1979,Scherer-REStat1982,Jaffe-AER1986}, or through flows of patent citations 
 \parencite[e.g.,][]{Jaffe-et-al-AER2000}.
Recent studies have explored more specific channels of knowledge spillovers. 
For example, \textcite{Bloom-et-al-ECTA2013} distinguished positive effects of R\&D spillover from negative ones in sharing product markets between firms.
\textcite{Zacchia-REStud2020} proposed a microeconomic model and showed evidence for knowledge spillover among firms through inventor networks.
Other studies have utilized exogenous termination of collaborations for the identification of peer effects \parencite[e.g.,][]{Waldinger-JPE2010
}
and positive externality from superstars in research \parencite[e.g.,][]{Azoulay-et-al-QJE2010}. 

Using patent data from Japan, this study contributes to the niche empirical literature by showing evidence for collaborative knowledge creation through 
active exchanges of knowledge among individual inventors, rather than passive improvement of productivity via spillovers.
Based on \textcite{Berliant-Fujita-IER2008}, our baseline regression model focuses on an inventor and their average collaborator.
It expresses the causality between their pairwise collaborative productivity and the differentiated knowledge of the average collaborator.
Collaborative output is measured in terms of quality and novelty of patents. 
For a given patent, the quality is evaluated by forward citations, and the novelty by the order of application within their primary technological category, reflecting the scarcity of invention in this category.

In this regression, we control for individual fixed effects by exploiting panel data and a variety of other factors. 
However, we face identification problems due to the remaining unobserved factors that influence inventors' collaboration and knowledge creation.
To deal with an endogenous regressor for an inventor (i.e., the differentiated knowledge of collaborators), we propose instrumental variables constructed from the same endogenous variables of their distant indirect collaborators.
Using more distant indirect collaborators to construct instrumental variables, one can reduce endogeneity caused by unobserved factors \parencite[e.g.,][]{Zacchia-REStud2020} and by reflection problems \parencite[e.g.,][]{Bramoulle-et-al-JE2009}, although this also makes the instruments weaker.
However, we argue that there is a channel in which our instruments retain relevance through firm-specific factors while maintaining their exogeneity.

Our baseline results indicate that a 10\% increase in collaborators' differentiated knowledge for an inventor raises the quality and novelty of their average pairwise output by 3\%--4\% and 5\%, respectively.
Moreover, we found that in the contribution of knowledge exchange to the quality and novelty of collaborative output, 17\% and 65\% can be attributed to the intensive margin (i.e., the average quality and novelty of output, respectively), rather than to the extensive margin (i.e., the number of patents developed).
Thus, collaborations appear to be more effective for seeking novelty of output.

The remainder of this paper is structured as follows. Section \ref{sec:bf} introduces the model by \textcite{Berliant-Fujita-IER2008}, and Section \ref{sec:fact} explains the motivating 
fact. Section \ref{sec:data} describes the data, Section \ref{sec:regression} details the regression models, and Section \ref{sec:iv} discusses our identification strategy. 
Section \ref{sec:results} presents the empirical results.
Finally, Section \ref{sec:conclusion} concludes the paper.

\section{A Mechanism of Collaborative Knowledge Creation\label{sec:bf}}

\noindent Here, we describe the Berliant-Fujita (BF) model. Each agent develops new knowledge either in isolation or by collaborating in pairs, building on their accumulated stock of knowledge. 
Consider a given set of agents who engage in knowledge creation.
They are assumed to be a priori homogeneous but become heterogeneous in the composition of the set of knowledge they created in the past.
Let $\delta_{ij}\in [0,1]$ be the proportion of time that agent $i$ allocates for collaboration with $j$.
If agent $i$ works in isolation, then their knowledge creation is subject to constant returns technology as given by $y_{ii}	 =\delta_{ii} a k_{ii}$ if $\delta_{ii}>0$ and 0 otherwise, where $a>0$, $k_{ii}$ is the knowledge stock of agent $i$, and $y_{ii}$ is the output.

If the subject instead collaborates with agent $j\,(\neq i)$, then their joint output is given by
\begin{equation}
y_{ij}	 = \delta_{ij} b \big(k_{ij}^C\big)^\theta \big(k_{ij}^D\big)^\frac{1-\theta}{2}\big(k_{ji}^D\big)^\frac{1-\theta}{2}\label{eq:bf-model}
\end{equation}
for $\delta_{ij}>0$ and $y_{ij}= 0$ otherwise, where $b>0$, $k_{ij}^C$ is the common knowledge of $i$ and $j$; $k_{ij}^D$ is the knowledge of agent $i$ differentiated from that of $j$; and $\theta\in (0,1)$ is the relative importance of common knowledge.
All pieces of knowledge (irrespective of timing at which they are created) are horizontally and symmetrically different.%
\footnote{The BF model of knowledge creation by exchanging mutual differentiated knowledge is consistent with the knowledge creation by recombination considered by \textcite{Weitzman-QJE1998} and \textcite{Olsson-JEG2000,Olsson-JEG2005}.}\

The output from the collaboration of agents $i$ and $j$ becomes their common knowledge.
Thus, the common knowledge of $i$ and $j$ increases as their collaboration lasts longer, and at the same time the differentiated knowledge between $i$ with agents other than $j$ also increases relative to their common knowledge.
To achieve the best balance between common and differentiated knowledge with collaborators, agents collectively decide on the group of collaborators, where each agent $i$ optimally chooses $\delta_{ij}$ for each $j$ of their collaborators to maximize the total output $\sum_{j} y_{ij}/2$ (assuming an equal split of output between collaborators).

Agents are assumed to maximize the present value of their lifetime knowledge production.
In a typical steady-state equilibrium, the size of the network component of each agent is given by $1+1/\theta$.%
\footnote{Myopic core is adopted as the equilibrium concept. A typical steady-state equilibrium is the one resulting from the initial state in which agents have sufficient common knowledge for the starting collaboration.}
Agents continue to rotate their collaborators so that the share $\delta = 1/(1+1/\theta)$ of the total time is allocated to each pairwise collaboration \parencite[][Proposition 1]{Berliant-Fujita-IER2008}.

While the model is highly stylized, it formalizes a microeconomic mechanism in which explicit knowledge exchange induces the creation of new knowledge.

\section{A Motivating Fact\label{sec:fact}}
\noindent Here, we define the key variables and present a motivating fact for our study.
We use patent application data from Japan and construct two-period balanced panel data by aggregating five consecutive years from 2000 to 2004 and from 2005 to 2009 to form periods 1 and 2, respectively.

Consider patent development in period $t\in \{1,2\}$ by the set $I$ of inventors in the panel and their collaborators. Let $\mathcal{G}_{it}$ be the set of patents in which inventor $i\in I$ participates in the development, and $G_j$ for $j\in \mathcal{G}_{it}$ is the set of inventors who participate in patent $j$.
Denoting the value of patent $j$ by a scalar $g_j>0$, the productivity of inventor $i$ is defined in terms of the total value of patents in which they participated, with the value of each patent being discounted by the number of inventors involved in the patent:
\begin{equation}
	\bar{y}_{it} = \sum_{j\in\mathcal{G}_{it}} \frac{g_j}{|G_j|}\,\label{eq:quality}
\end{equation}
where $|G_j|$ is the cardinality of set $G_j$. Hereafter, the expression $|X|$ for a set $X$ denotes the cardinality of $X$.
Let
\begin{equation}
	N_{it} \equiv \cup_{j\in \mathcal{G}_{it}} G_j\backslash\{i\} \label{eq:n}
\end{equation}
represent the set of collaborators of inventor $i\in I$ such that each agent in $N_{it}$ participates in at least one common patent with $i$ in period $t$. 
Then,
\begin{equation}
    y_{it}\equiv \frac{\bar{y}_{it}}{n_{it}}
\end{equation}
represents the \emph{average pairwise output} by inventor $i$ in period $t$, where $n_{it}\equiv |N_{it}|$ is the total number of collaborators of $i$ in period $t$. It can be interpreted as the collaborative output of inventor $i$ with their average collaborator, corresponding to $y_{ij}$ in \eqref{eq:bf-model}.

We focus on the role of knowledge exchange in the BF model, and hence the differentiated knowledge $k_{ji}^D$ of collaborators in \eqref{eq:bf-model}.
To simplify the analysis, we assume that the cumulative knowledge of an inventor is fully utilized in the knowledge creation in each period, and hence it is reflected in their output in the same period.
Accordingly, \emph{the average differentiated knowledge of collaborators} of inventor $i$ in period $t$ is assumed to be reflected in the average output that the collaborators of $i$ produced outside the joint projects with $i$ in the same period:
\begin{equation}
	k^{D}_{it} =\frac{1}{n_{it}}\sum_{j\in N_{it}}\: \sum_{k\in \mathcal{G}_{jt}\backslash\mathcal{G}_{it}} \frac{g_k}{|G_k|}.\label{eq:collaborator-productivity}
\end{equation}

Turning to the quantification of collaborative productivity, we consider the quality and novelty dimensions. 
The quality of a patent is measured by cited counts. Specifically, $g_{j}$ represents the count of citations that patent $j$ received in five years after publication.
Here, $g_{j}$ excludes citations from the patents in which any firm employing the inventors in $G_j$ is involved.%
\footnote{More precisely, let period 3 include all years from 2010, and let $\mathcal{G}$ be the set of all patents applied in periods $t = 1,2$ and $3$.
Let $F_{it}$ be the set of inventors who belong to the same firm as inventor $i$ in period $t$. Then, the set of potential citing patents for patent $j$ is given by $C_j \equiv \{k\in \mathcal{G}: k\notin \mathcal{G}_{\ell t}\:\forall\, \ell\in F_{it} \:\forall\, i\in G_j\: \forall\, t \in \{ 1, 2,3\}\}$. Note that our quality-measure is far more conservative than a simple exclusion of self-citation at the inventor or firm level.}\ 
This construction eliminates obvious correlations between $y_{it}$ and $k_{it}^D$ due to mutual citations among the firms affiliating in R\&D.

The novelty of a patent, on the other hand, is measured in terms of the scarcity of existing patents sharing the same technological category with the patent. Specifically, $g_j$ is defined by the reciprocal, $1/r_j$, of the order, $r_j = 1,2, \ldots$, of $j$ with respect to its application date among all the patents classified in the same technological category as $j$.
In our analysis, the technological category adopted to compute the patent novelty is a `subgroup' of the International Patent Classification (IPC) and is the most disaggregated technological category available in the data.
There were 31,511 and 26,424 subgroups with positive numbers of patents applied in periods 1 and 2, respectively.%
\footnote{Refer to Appendix \ref{app:data} (Table \ref{tb:descriptive-stat}) as well as Appendix \ref{app:ipc} for more details on technological categories in IPC.}\ 

Alternatively, the novelty of a patent has been defined, for example, by \textcite{Fleming-et-al-ASQ2007} to be the number of new subclass pairs associated with the patent.%
\footnote{They used the United States Patent Classification with ~10,000 subclasses.}\ 
However, this measure is too extreme for our purpose, since a patent's novelty is zero unless it has at least a new subclass combination.
A recent measure of novelty by \textcite{Watzinger-Schnitzer-CEPR2019} is based on the frequency of word combinations appearing in the text of a patent in older patents. It may capture some aspects that are missed in our definition of novelty, and vice versa. However, the correspondence between arbitrary word combinations and technologies is not clear at this point.
We thus adopt a novelty measure that is consistent with the technological categories of IPC, which
defines the similarity and dissimilarity of all the patents in our data.

\begin{figure}[h!]
\includegraphics[scale=.1333]{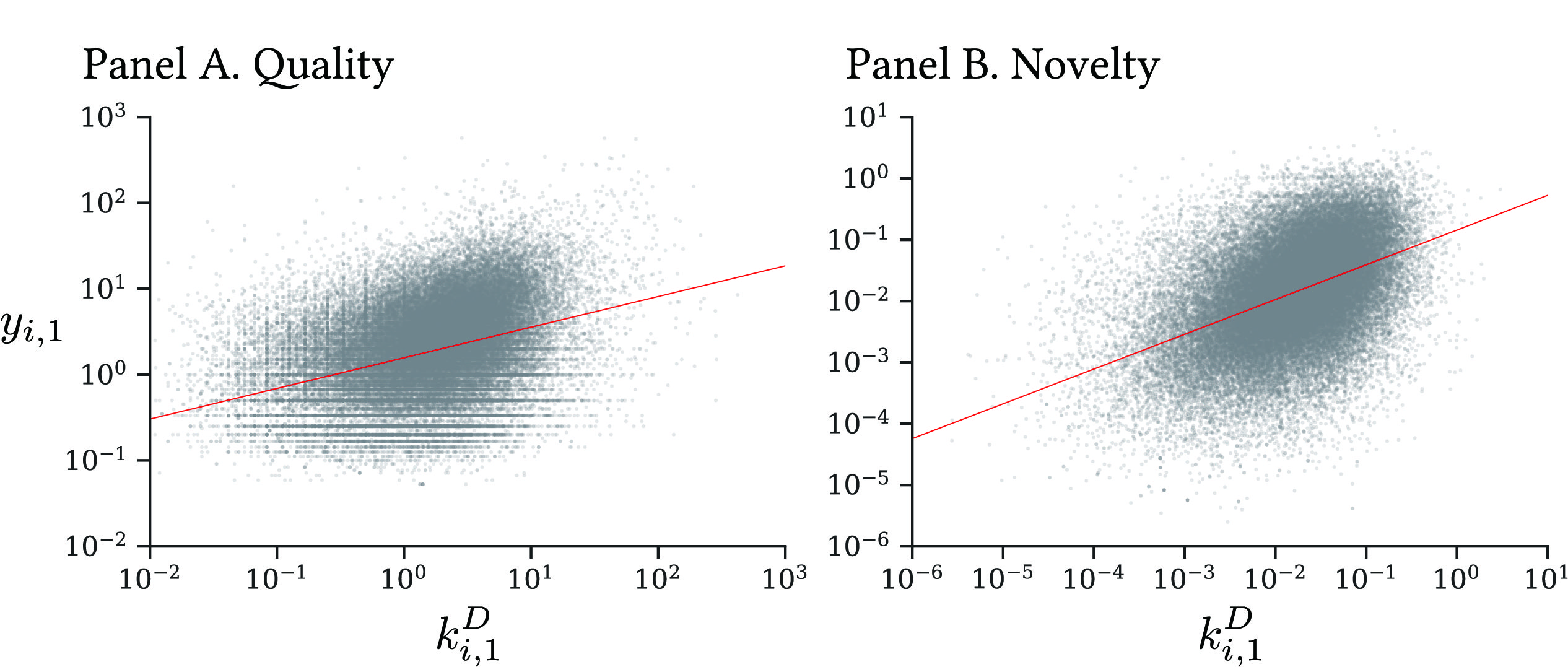}
\caption{Pairwise productivity versus pairwise differentiated knowledge of inventors\label{fig:fact}}
\begin{figurenotes}[\fontsize{10pt}{10pt}\selectfont Notes]
\fontsize{10pt}{10pt}\selectfont
The figure plots the average pairwise productivity of an inventor $y_{i,1}$ against the average differentiated knowledge of their collaborators $k_{i,1}^D$ for period 1 in log-scale. Productivity is measured based on the quality and novelty of patents for which the inventor participated in the development of. 
The simple ordinary least squares (OLS) regression line is shown in each panel. The estimated coefficients of $\ln k_{it}^D$ are 0.357 and 0.567 in Panels A and B, respectively.
\end{figurenotes}
\end{figure}

Figure \ref{fig:fact} shows the relation between $y_{i1}$ and $k_{i1}^D$ in period 1. 
Panels A and B indicate a clear positive relationship between the two variables under quality and novelty measures, respectively.
In particular, for the quality-based productivity in Panel A, by construction, the positive correlation does not accrue from mutual citations among collaborators or their employer firms.
This provides us a motivating fact for investigating 
the causal effect suggested by the BF model.
%

\section{Data\label{sec:data}}

\noindent {\it Patent data} -- Patent data are taken from the \emph{published patent applications} of Japan \parencite[][]{Naito-Data}, including information on all patents applied for between 1993 and 2017.
Since every applied patent is published within 1.5 years of application, our data include all the output of inventors at a given point in time during the study period.

In this data, identical inventors are traced by matching their names and the establishments they belong to, recorded in the applications of patents in which they participated.
By utilizing this information, we focus on the panel of ${\rm 29,287}$ inventors constituting set $I$ who had been active and stayed in the same establishments (and hence the same firms) in both periods, and they have up to the 5th indirect collaborators in each period. This panel helps us to isolate inventor and firm-specific factors that may influence knowledge creation.
The indirect collaborators are used to construct instrumental variables in order to eliminate influence from other endogenous factors.
Descriptive statistics of the data concerning the inventors in $I$ are shown in Appendix \ref{app:data}.

The information on inventors not included in $I$ is still used if they either directly or indirectly collaborate with inventors in $I$. Specifically, there are 434,555 and 283,860 direct and indirect collaborators in total in periods 1 and 2, respectively.
The inventors in $I$ for regressions account for only 6.7\% and 8.8\% of all the inventors with at least one collaborator in periods 1 and 2, respectively.
Nonetheless, neither the quality nor novelty of the developed patents exhibit stark differences between the two sets of inventors.
For the average quality (novelty) of patents in pairwise collaborative output, $\ln y_{i,t}/|G_{i,t}|$ in period 1, the mean values are -3.95 and -3.02 (-8.82 and -8.61) with standard deviations of 1.77 and 1.97 (2.01 and 2.61) for inventors in $I$ and for the full samples, respectively (Figure \ref{fig:productivity_dist} in Appendix \ref{app:data} compares their distributions between the selected and full samples).

Each patent is associated with at least one technological classification based on the IPC, which is maintained by the World Intellectual Property Organization (\url{http://www.wipo.int/portal/en/index.html}). 
The IPC hierarchically classifies technologies into eight sections, ~120 classes, 300 subclasses, and finally ~40,000 subgroups. 
The IPC's labeling scheme is consistent over time, and a newly introduced category is basically associated with a new technology. 
Hence, the classifications in IPC at a given point in time roughly represent the state-of-the-art technological categories at that time.%
\footnote{See Appendix \ref{app:ipc} for the details of the IPC technological categories.}\ 
Although an applicant can claim more than one IPC category for their patent, we adopted only the primary IPC category of each patent to represent its technological category to avoid subjective variation.
We adopt the finest subgroup classification to define the novelty of the patents, which together comprise 40,691 and 38,339 categories associated with the patents in our data in periods 1 and 2, respectively.

\smallskip
\noindent {\it Urban agglomerations} -- R\&D activities are disproportionately concentrated in large cities (see Figure \ref{fig:ua} in Appendix \ref{app:locational-factors}).
If an \emph{urban agglomeration} (UA) is defined as a contiguous area of population density of at least 1000/km$^2$ and with a total population of at least 10,000,%
\footnote{Population data are obtained from the Population Census (\citeyear{Pop-Census-2010}) by MIAC.}\ 
in 2000, 99\% of all inventors concentrated in UAs, with 81\% in the largest three UAs.
Inventors located within a 10 km buffer of any of the 453 UAs are assigned to the closest UA; otherwise, their locations are considered to be rural. In the regressions, the standard errors are clustered by UAs.%
\footnote{As UAs spatially expand over time on average, we use the boundaries of UAs in 2010, each of which provides the largest spatial extent during the study period 1995--2009 on average. However, the choice of the particular time point should not affect the basic results because most inventors are concentrated in relatively large UAs whose spatial coverage is relatively stable over the study period.}\ 

\smallskip
\noindent {\it Geographic neighborhood factors} -- Given the disproportionate geographic concentration of R\&D activities, productivity is expected to be influenced by geographic neighborhood \parencite[e.g.,][]{Jaffe-Trajtenberg-Henderson-QJE1993,Thompson-Fox-Kean-2005,Kerr-Kominers-REStat2015}. We control for the geographic concentrations of five types of activities: inventors, R\&D expenditure, manufacturing employment and output, and residential population.
Each geographic concentration is defined by the size of the concentration in a circle of given radius around inventor $i$. The formal definitions and descriptive statistics are described in Appendix \ref{app:locational-factors}.

\section{Regression Models\label{sec:regression}}
\noindent Here, we describe the regression models.
We focus on collaborative cases and do not address the possibility of working in isolation. 
To apply data to theory, the original specification is simplified by 
formulating a regression model for knowledge creation between an inventor and their average collaborator rather than between each pair of inventors:
\begin{equation}
	\ln y_{it} = \alpha + \beta \ln k_{it}^D + \gamma_1 \ln k_{it} + \gamma_2 \left(\ln k_{it}\right)^2 + \ln A_{it} + \lambda_i + \tau_t + \varepsilon_{it}\,. \label{eq:baseline}
\end{equation}
In the theoretical model, we focus on knowledge exchange, and hence the role of the differentiated knowledge $k_{ji}^D$ of collaborators in \eqref{eq:bf-model}, while abstracting from the role of common knowledge $k_{ij}^C$ and differentiated knowledge $k_{ij}^{D}$ of inventor $i$.
The effect of knowledge exchange should naturally be reflected in that of the average differentiated knowledge of collaborators, $\ln k_{it}^D$ in \eqref{eq:baseline}.

To control for other non-negligible knowledge effects, we include the cumulative research scope of inventor $i$s’ past projects.
For this purpose, let $S$ denote the set of all technological categories (i.e., IPC subgroups), and the technological category assigned to patent $j$ be $s_j\in S$. 
Then, the research scope of inventor $i$ in period $t$ is defined by
\begin{equation}
	S_{it} = \cup_{j\in \mathcal{G}_{it}} \{  s_j \}\label{eq:specialization}
\end{equation}
which in turn can be used to define the \emph{cumulative research scope} of inventor $i$ in period $t$ by%
\footnote{To compute $k_{it}$ for $t\in\{1,2\}$, we use all available data from 1993. Specifically, we define period 0 to be from 1993 to 1999.}\ 
\begin{equation}
	k_{it} =\Big|\underset{t'<t}{\cup} S_{it'}\Big|\,.\label{eq:research-scope}
\end{equation}
The past experience of inventor $i$ reflected in this value is naturally expected to correlate with the size of common knowledge between inventor $i$ and their collaborators as well as the differentiated knowledge of $i$.
In addition, it may partly control for other factors discussed in the literature; for example, technological obsolescence \parencite[e.g.,][]{Horii-JEDC2012}, imitations \parencite[e.g.,][]{Chu-JEG2009,Cozzi-Galli-JEG2014}, and learning-by-doing effects  \parencite[e.g.,][]{Grossman-Helpman-Book1991,Klette-Kortum-JPE2004, Lucas-Moll-JPE2014}. 
The third and fourth terms on the RHS of \eqref{eq:baseline} are supposed to capture their overall effects up to the second order.

Other inventor- and time-specific productivity shifters 
for inventor $i$ are bundled in the fifth term, $A_{it} \equiv e^{{\bm X}^\top_{it}{\bm\eta}}$. 
Namely, $\bm{X}_{it}$ represents a vector of geographic neighborhood effects defined in Section \ref{sec:data} that includes local concentrations of other inventors, R\&D expenditure, manufacturing employment/production, and residential population; $\bm \eta$ is a vector of coefficients corresponding to each element of $\bm X_{it}$.
The last three terms, $\lambda_i$, $\tau_t$, and $\varepsilon_{it}$, on the RHS are the time-invariant inventor fixed effect, period fixed effect, and inventor- and period-specific error, respectively.
The values of parameters $\alpha, \beta, \gamma_1, \gamma_2, \bm{\eta}$, and $\tau_t$ are estimated by regressions.

Note that the technology of knowledge creation \eqref{eq:bf-model} itself, and hence the corresponding empirical model \eqref{eq:baseline}, is not restricted to
any specific mechanism of inventors' collaboration, although the BF model assumes autonomous collaborations by inventors. 
Consequently, in \eqref{eq:baseline}
it does not matter for parameter identification whether network formation is autonomous or determined at the firm or establishment level. 

Finally, we exploit the log-linear relation between quantity and quality (or novelty) of collaborative productivity given by \eqref{eq:quality}:
\begin{equation}
	\ln y_{it} = \ln y_{it}^p + \ln y_{it}^q\,.\label{eq:decomposition}
\end{equation}
In the first term on the RHS of \eqref{eq:decomposition}, 
$y_{it}^p$ denotes the number of patents included (i.e., the extensive margin) in inventor $i$'s pairwise output given by $\bar{y}_{it}^p / n_{it}$, where $\bar{y}_{it}^p \equiv \sum_{j\in \mathcal{G}_{it}} 1/|G_j|$, which coincides with $\bar{y}_{it}$ under $g_j=1$ in \eqref{eq:quality}.
In the second term, $y_{it}^q$ represents the average quality or novelty (i.e., the intensive margin) of $i$'s pairwise output,   $y_{it}/y_{it}^p$.
We can thus decompose the effect of each explanatory variable in \eqref{eq:baseline} into those related to the quantity and average quality/novelty of inventors' pairwise output $y_{it}$. 
The model to be estimated for this purpose is given by
\begin{equation}
	\ln y_{it}^m = \alpha^m + \beta^m \ln k_{it}^D + \gamma_1^m \ln k_{it} + \gamma_2^m \left(\ln k_{it}\right)^2 + \ln A_{it}^m + \lambda_i^m + \tau_t^m + \varepsilon_{it}^m\, \label{eq:quality-equation}
\end{equation}
for $m \in \{p,q\}$, where the coefficients of each explanatory variable for $m$ add up to that of the corresponding variable in \eqref{eq:baseline}. 
In particular, we have $\beta = \beta^p + \beta^q$ for the coefficient of $\ln k_{it}^D$.

\section{Identification by Instrumental Variables\label{sec:iv}}
\noindent Here, we present our strategy for model identification by dealing with the endogeneity of the average differentiated knowledge of collaborators $\ln k_{it}^D$ in \eqref{eq:baseline}.
There are three sources of endogeneity. 
The first results from inventors' endogenous collaboration; that is, network endogeneity, where unobservable influences exist on inventors' collaboration decisions and their productivities \parencite[e.g.,][]{Goldsmith-Pinkham-Imbens-JBES2013}.
The second results from the mutual dependence of productivities between an inventor and their collaborators through $k_{it}^D$ in \eqref{eq:baseline}. This is the so-called `reflection problem' in the context of econometric network analysis \parencite[e.g.,][]{Manski-REStud1993, Bramoulle-et-al-JE2009}.
The third arises from unobservable network specific factors that influence an inventor and their collaborators’ productivities simultaneously. 
These are called `correlated effects' in \textcite{Manski-REStud1993}.
To solve the endogeneity caused by these reasons, we argue that the endogenous variables $k_{it}^D$ in \eqref{eq:baseline} for inventor $i$ can be instrumented by the average value of the same variable for the distant indirect collaborators of $i$.

Let $\bar{N}_{it}^\ell$ be the set of all the 0-th to $\ell$-th indirect collaborators of inventor $i$ given by
\begin{align}
	\bar{N}_{it}^\ell &= \bar{N}^{\ell-1}_{it} \raisebox{-0.2ex}{$\medcup$} \left[\raisebox{-0.2ex}{$\medcup$}_{j\in \bar{N}_{it}^{\ell-1}} N_{jt} \right]\,\quad \ell = 1,2,\ldots\label{eq:indirect-collaborators}
\end{align}
where the set of the `0-th indirect collaborators' is defined by the set of inventors comprising $i$ and their direct collaborators $\bar{N}_{it}^0 \equiv N_{it} \cup \{i\}$.
To obtain $\bar{N}^\ell_{it}$ from $\bar{N}^{\ell-1}_{it}$ for $\ell=1,2,\ldots$, we expand $\bar{N}^{\ell-1}_{it}$ by the union of all the direct collaborators of $j\in \bar{N}^{\ell-1}_{it}$ as in \eqref{eq:indirect-collaborators}.
The set of the \emph{$\ell$-th indirect collaborators} of $i$ can then be given by
\begin{align}
	N_{it}^\ell &= \bar{N}_{it}^\ell \backslash \bar{N}_{it}^{\ell-1}\, \quad \ell= 1,2, \ldots
\end{align}
The instrument $k_{it}^{D,\text{IV}_\ell}$ for $k_{it}^D$ can be constructed as the average values of the differentiated knowledge of collaborators for $\ell$-th indirect collaborators $j \in N_{it}^\ell$:
\begin{align}
	k_{it}^{D,\text{IV}_\ell} = \frac{1}{n_{it}^\ell} \sum_{j\in N_{it}^\ell}  k_{jt}^D\label{eq:iv}
\end{align}
where $n_{it}^\ell \equiv \left| N_{it}^\ell\right|$. 
\medskip

\noindent {\it Exogeneity of the instruments} --
The extant literature on social interactions \parencite[e.g.,][]{Bramoulle-et-al-JE2009, De-Giorgi-et-al-AEJ2010, Calvo-Armengol-et-al-REStud2009} suggests that the reflection problem in our context be reduced by using instruments constructed from indirect collaborators. Namely, the farther an indirect collaborator is from an inventor in the collaboration network, the smaller the influence of their output on the inventor's productivity.%
\footnote{For example, in eq.\,(6) of \textcite{Bramoulle-et-al-JE2009}, the endogenous peer effect from the $\ell$-th indirect peer is given by $\beta^{1+\ell}$, where $\beta\in (0,1)$ and $\ell=0,1,2,\ldots$ with the $0$-th indirect peer being the direct peer. The peer effect $\beta^{1+\ell}$ from the $\ell$-th indirect peers diminishes as $\ell$ increases.}\ 
The instruments further solve the endogeneity caused by the endogenous network and correlated effects, if more distant inventors share less unobserved factors.

This effect applies to our case if, for example, inventors with similar (observable and unobservable) characteristics have proclivities to collaborate with each other.
An obvious situation is that inventors belong to the same firm. However, this case is not an issue for us, since firm-specific factors can be controlled by inventor fixed effects.
Another typical situation is that inventors have similar technological specializations. These inventors likely share opportunities to exchange ideas with each other through, for example, conferences and journals of common research subjects, thereby affecting their chance of collaboration and productivities.
The exogeneity of the instruments in this case arises from the fact that more distant indirect collaborators share less common research fields with each other.\footnote{Our logic for the exogeneity of the instruments is similar to that of \textcite{Zacchia-REStud2020}, who used similar instruments to identify knowledge spillovers between firms through their inventor networks.}

To see if our data supports this conjecture, we quantify the commonality in technological specialization between indirect collaborators using the Jaccar index.
The average Jaccar index between the technological specialization $S_{it}$ of inventor $i$ and those of their $\ell$-th indirect collaborators $j\in N_{it}^\ell$ is computed as:
\begin{equation}
	\mathrm{J}^\ell_{it} = \frac{1}{n_{it}^\ell} \sum_{j \in N_{it}^\ell} \frac{|S_{it} \cap S_{jt}|}{|S_{it} \cup S_{jt}| }\in [0,1]\,.
\end{equation}
A larger value of $\mathrm{J}^\ell_{it}$ implies higher average similarity in technological specialization between inventor $i$ and their $\ell$-th indirect collaborators. 
In particular, it takes value 0 if their specializations do not overlap (i.e., $S_{it}\cap S_{jt} = 0\:\forall j\in N_{it}^\ell$) and value 1 if they are identical (i.e., $S_{it} = S_{jt}\:\forall j\in N_{it}^\ell$).

\begin{figure}[h!]
\includegraphics[scale=.17]{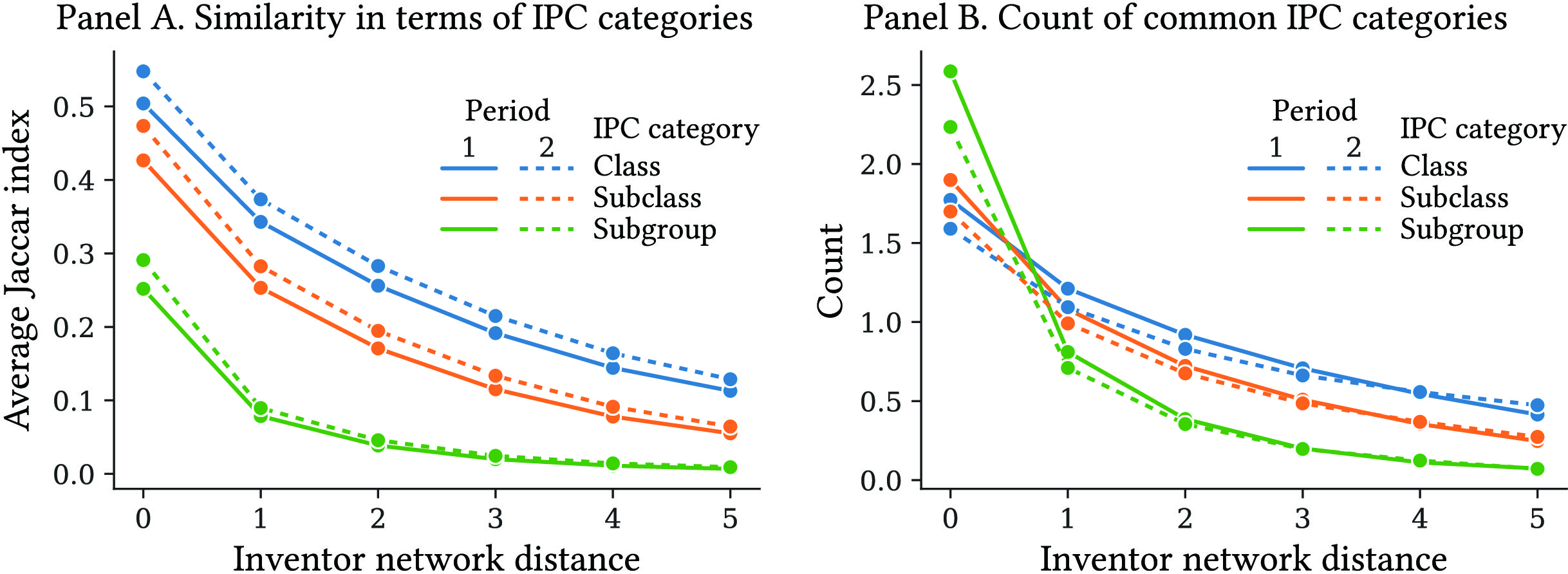}
\caption{Exogeneity of instrumental variables\label{fig:exogeneity}}
\begin{figurenotes}[\fontsize{10pt}{10pt}\selectfont Notes]
\fontsize{10pt}{10pt}\selectfont
Panel A plots the average value of Jacaar indices between the research scopes $S_{it}$ of direct and indirect collaborators of inventors for $t=1,2$, where the research scope is defined in terms of the IPC categories. The horizontal axis indicates the degrees of indirectness in the collaborator network, where 0 represents a direct collaborator, and values of $j=1, 2,\ldots,5$ represent the $j$-th indirect collaborators.
Panel B plots the average counts of common IPC categories between an inventor and direct/indirect collaborators.
\end{figurenotes}
\end{figure}

In Figure \ref{fig:exogeneity}, Panel A depicts the average values of $\mathrm{J}^\ell_{i1}$ overall $i\in I$ in terms of IPC sections, classes, subclasses, and subgroups between an inventor and their $\ell$-th indirect collaborators for $\ell = 0,1,\ldots,5$.
Panel B complements it by showing the average count of common technological categories in the corresponding classifications between an inventor and their $\ell$-th indirect collaborators.
From Panel A, the commonality of specializations steadily decreases as the indirectness $\ell$ increases. 
In terms of the IPC subgroup, it almost vanishes for the 3rd indirect collaborators. 
In terms of the average count of common categories, it is less than 1 for the 3rd and higher indirect collaborators for all classifications in IPC.

\smallskip

\noindent {\it Relevance of instruments} -- Suppose that endogeneity is partly caused by some time-invariant unobserved factors specific to the firms or establishments that inventors belong to.
In this situation, there is a possible channel over the collaboration network through which our instruments retain relevance while satisfying exogeneity.
We elaborate a simple example here.
Note that the set $N_{it}$ of collaborators and the set $N_{it}^\ell$ of $\ell$-th indirect collaborators of inventor $i$ may contain inventors that belong to the same firms, but which are different from the one $i$ belongs to.
For these inventors, the productivities, and hence $k_{it}^D$ and $k_{it}^{D,\rm{IV}_\ell}$, correlate through the firm-specific factors.%
\footnote{Alternatively, $k_{it}^D$ and $k_{it}^{D,\rm{IV}_\ell}$ also correlate if the 1st-indirect and $\ell$-th indirect collaborators of $i$ belong to the same firm.}\ 
On the other hand, they do not have common firm-specific factors with $i$.
Consequently, there is no correlation between $y_{it}$ and $k_{it}^{D,\rm{IV}_\ell}$ through unobserved factors.%
\footnote{Although $N_{it}^\ell$ possibly contains inventors that belong to the same firms as $i$, their common firm-specific factors are controlled by inventor $i$'s fixed effect and hence do not violate the exogeneity condition.}\ 

Whether there is sufficiently strong relevance of instruments is an empirical question to be examined in Section \ref{sec:results}.
The success in our case hinges on the fact that a relatively large share of distant indirect collaborators belong to the same firms. 
Specifically, the shares of the 3rd, 4th, and 5th-indirect collaborators of an inventor who belong to the same firm as the inventor are 50\%, 38\%, and 25\% (54\%, 43\%, and 32\%) on average, respectively, in period 1 (period 2).

\section{Results\label{sec:results}}
\noindent Here, we present our main regression results for models \eqref{eq:baseline} and \eqref{eq:quality-equation} with brief discussions on robustness checks, the details of which are given in Appendix \ref{app:robustness}.
In all of the regressions conducted, the fixed effects of inventors, periods, and IPC classes are controlled.
The geographic neighborhood factors described in Section \ref{sec:data} are constructed for a circle with a 1 km radius around each inventor except for residential population, which is computed within a 20 km radius around an inventor to account for the urban environment around them.
Standard errors are clustered by UAs (see Section \ref{sec:data}).%
\footnote{As the instruments $\ln k_{it}^{D, \text{IV}_\ell}$ for $\ln k_{it}^D$ in \eqref{eq:baseline} and \eqref{eq:quality-equation} involve inventors located in different UAs, one might suspect that cluster-robust standard errors are incorrect because the instruments for any inventor $i$ might be correlated with errors $\varepsilon_{jt}$ in \eqref{eq:baseline} for any inventor $j$ even if inventors $i$ and $j$ are located in different UAs. 
However, we consider that these cluster-robust standard errors still provide correct standard errors because the inventor fixed effects controlled in all regressions encompass UA-specific fixed effects, making the errors free from correlation with UAs while allowing for standard errors to vary across UAs.
}\

\medskip

\noindent {\it Baseline results} -- Panels A and B in Table \ref{tb:bf-model} summarize the regression results for key variables of model \eqref{eq:baseline} under quality- and novelty-adjusted productivity, respectively.
More detailed tables with estimated coefficients for neighborhood factors are described in Appendix \ref{app:baseine} (Tables \ref{tb:bf-model-citation-full-result} and \ref{tb:bf-model-novelty-full-result}).%
\footnote{The estimated coefficients of the key variables in Table \ref{tb:bf-model} change only marginally under alternative radius values (5 and 10 km) adopted to compute geographic neighborhood factors.}\ 
In both panels, column 1 reports the result from the OLS regression, and the rest report those from two-stage least squares (2SLS) instrumental variable (IV) regressions.
For the IV regressions, we use the 3rd to the 5th indirect collaborators to construct IVs for $\ln k_{it}^D$. 
More specifically, we use all three instruments $\ln k_{it}^{\text{IV}_\ell}$ for $\ell = 3,4$ and $5$ in column 2, while only one is used in columns 3--5, respectively.

The IV results support the role of knowledge exchange in the BF model.
The estimated coefficients of $\ln k_{it}^D$ are persistently positive and similar, 0.33--0.39 and 0.48--0.51, for quality- and novelty-based productivity, respectively (except for the IV5-result in Panel A, which is discussed later). The values below 1 indicate decreasing returns to knowledge exchange, which is consistent with the implication of the BF model that the benefit from collaborators' differentiated knowledge will eventually be dominated by that of common knowledge with collaborators and 
the inventors' own differentiated knowledge. 

The estimated positive effect of research scope $\ln k_{it}$ of an inventor and the negative effect of its squared term $(\ln k_{it})^2$ are consistent with the interpretation of positive decreasing-returns effects of common knowledge
as well as differentiated knowledge
of the focal inventor, together with other possible effects such as learning-by-doing effects discussed in Section \ref{sec:regression}.


To examine the strength of instruments, we conducted the heteroskedasticity robust weak instrument test of \textcite{Olea-Pflueger-JBES2013}.
Except for the IV regression based on the 5th indirect collaborators for the quality-based case (column 5 in Panel A), 
\textcite{Olea-Pflueger-JBES2013}'s first-stage effective $F$ statistics, $F^{\rm{eff}}$, take large values, meaning that the IVs do not seem to be weak.%
\footnote{See Table \ref{tb:first-stage} in Appendix \ref{app:baseine} for the results of the first-stage regressions.}\ 
Indeed, the estimated coefficient of $\ln k_{it}^{D}$ in column 5 in Panel A (which is instrumented by $\ln k_{it}^{D,\rm{IV}_5}$) seems to differ from the others.
To confirm the exogeneity of the IVs, we use $\ln k_{it}^{\text{IV}_\ell}$ for all $\ell = 3,4$ and $5$ in column 2 and conduct Hansen's (\citeyear{Hansen-ECTA1982}) $J$ test for overidentifying restrictions.
The \emph{p}-values of the test are 0.68 and 0.11 for productivities based on quality and novelty, respectively, meaning that the exogeneity of the IVs cannot be rejected.%
\footnote{\label{ft:hansen}Of course, this result of Hansen's $J$ test is by no means sufficient to guarantee the exogeneity of the instruments if all the instruments are subject to the same type and magnitude of bias.}\ 

The OLS result is consistent with the IV results in terms of the signs of the estimated coefficients, but it appears to have a substantial downward bias in the estimated coefficient of $\ln k_{it}^D$ under both measures of productivity.%
\footnote{\textcite{Akcigit-et-al-NBER2018} and \textcite{Zacchia-REStud2020} reported similar downward bias on the effects of spillovers/interactions from other agents on the R\&D outcome.}\ 
A possible explanation for the bias is that the more productive inventors attract (or are assigned by their firm) a larger number of relatively unexperienced collaborators than the inventor intends. The removal of this reverse causality leads to a larger positive effect in the IV estimates.
\begin{table}[p]
\centering
\caption{Regression results for the Berliant-Fujita model\label{tb:bf-model}}
\begin{tabular*}{\textwidth}{p{1in}@{\extracolsep{\fill}}ccccc} 
\midrule[0.6pt]\midrule[0.6pt]
& \multicolumn{5}{c@{}}{Panel A. Quality} \\[2pt]
\cmidrule[0.6pt]{2-6}
Variable &(1) OLS & (2) IV3-5 & (3) IV3 & (4) IV4 & (5) IV5\\
\midrule[0.6pt]
$\ln k_{it}^D$ & 0.166 & 0.334 & 0.340 & 0.392 & 0.488 \\[-2pt]
& (0.0111) & (0.0444) & (0.0440) & (0.0565) & (0.118) \\[2.5pt]
$\ln k_{it}$ & 0.163 & 0.140 & 0.139 & 0.132 & 0.119 \\[-2pt]
& (0.0326) & (0.0213) & (0.0212) & (0.0199) & (0.0192) \\[2.5pt]
$(\ln k_{it})^2$ & -0.0744 & -0.0669 & -0.0666 & -0.0643 & -0.0600 \\[-2pt]
 & (0.0116) & (0.00818) & (0.00816) & (0.00765) & (0.00721) \\
 \cmidrule[0.6pt]{1-6}
$R^2$ & 0.123 & 0.106 & 0.104 & 0.091 & 0.059 \\
$F^{\mbox{eff}}$  & & 52.96 & 179 & 41.90 & 8.413 \\
\multicolumn{2}{l@{}}{Critical $F^{\mbox{eff}}$-value}  & 20.09 & 23.11 & 23.11 & 23.11 \\ 
\multicolumn{2}{l@{}}{Hansen $J$ $p$-value}   & 0.681 &  &  &  \\
\midrule[0.6pt]
\\
& \multicolumn{5}{c@{}}{Panel B. Novelty} \\[2pt]
\cmidrule[0.6pt]{2-6}
Variable & (1) OLS & (2) IV3-5 & (3) IV3 & (4) IV4 & (5) IV5\\
\midrule[0.6pt]
$\ln k_{it}^D$ & 0.217 & 0.480 & 0.478 & 0.511 & 0.495 \\[-2pt]
& (0.00693) & (0.0403) & (0.0409) & (0.0335) & (0.0491) \\
$\ln k_{it}$& 0.235 & 0.189 & 0.190 & 0.184 & 0.187 \\[-2pt]
& (0.0187) & (0.0166) & (0.0166) & (0.0174) & (0.0202) \\
$(\ln k_{it})^2$ & -0.182 & -0.161 & -0.161 & -0.158 & -0.160 \\[-2pt]
& (0.0148) & (0.00924) & (0.00920) & (0.00982) & (0.00916) \\
 \cmidrule[0.6pt]{1-6}
$R^2$& 0.175 & 0.134 & 0.135 & 0.124 & 0.129 \\
$F^{\mbox{eff}}$  && 332.9 & 1009 & 720.8 &  166.3 \\
\multicolumn{2}{l@{}}{Critical $F^{\mbox{eff}}$-value } & 17.85 & 23.11 & 23.11 & 23.11 \\
\multicolumn{2}{l@{}}{Hansen $J$ $p$-value} & 0.113 &  &  &  \\
\midrule[0.7pt]
\end{tabular*}
\begin{tablenotes}[\fontsize{10pt}{10pt}\selectfont Notes]
\fontsize{10pt}{10pt}\selectfont
(i) The number of observations is 58,574 (29,287 inventors$\times$2 time periods) for all regressions. The inventors included have up to at least the 5th indirect collaborators in order to construct instruments for an endogenous variable. 
(ii) The dependent variable is the log of the average pairwise productivity of an inventor, $\ln y_{it}$. It is defined in terms of cited counts and novelty of patents in Panels A and B, respectively. 
(iii) The explanatory variables shown are the average differentiated knowledge of collaborators, $\ln k_{it}^D$; the first- and second-order effects of the research scope, $\ln k_{it}$, of an inventor.
(iv) Column 1 shows the result from the OLS regression, while columns 2--5 show the results of IV regressions, where $\ln k_{it}^D$ is treated as an endogenous variable, and is instrumented by the same variable of the 3rd-5th, 3rd, 4th, and 5th indirect collaborators of inventor $i$ in columns 2, 3, 4, and 5, respectively. 
(v) In all of the regressions, the inventor, year, and IPC class fixed effects, as well as a variety of neighborhood effects are controlled. Neighborhood effects include the sizes of concentration of inventors, R\&D expenditure, manufacturing employment and output within a 1-km radius, and population within a 20-km radius around a given inventor. 
(vi) The fifth row in each panel reports \textcite{Olea-Pflueger-JBES2013}'s effective first-stage $F$-statistic.
(vii) The sixth row in each panel indicates the 5\% critical value of the effective $F$-statistic with a Nagar bias threshold $\tau=10\%$. 
(vii) The last row in each panel indicates the $p$-value of \textcite{Hansen-ECTA1982}'s $J$ test.
(viii) Robust standard errors in parentheses are clustered on urban agglomeration.
\end{tablenotes}
\end{table} 

\medskip

\noindent {\it Robustness checks} -- The estimated coefficient of $\ln k_{it}^D$ in \eqref{eq:baseline} might reflect not only the effects of differentiated knowledge of collaborators, but also those of time-varying factors specific to the inventors' own and collaborators' firms or establishments. Such factors include the R\&D environment and productivity externality (peer effects) from (possibly non-collaborating) inventors that vary across firms or establishments, and their research affiliations.

To investigate these possibilities, we conduct the two exercises (see Appendix \ref{app:robustness} for details).
First, we include the size and research scope of the firm or establishment to which an inventor belongs as additional explanatory variables in \eqref{eq:baseline}. Second, we consider possible influences from the broader neighborhood in the research network beyond an inventor's own firm or establishment. 
Specifically, we simulate random counterfactual choices of collaborators for each inventor in $I$ conditional on the actual number of collaborators as well as the firms/establishments to which these collaborators belong.
We then replace $k_{it}^D$ in \eqref{eq:baseline} with that constructed from the counterfactual collaborators and estimate the model.
We find that the effect of knowledge exchange on the collaborative productivity appears to be at most mildly influenced by factors at the level of an establishment, a firm, and a research affiliation of firms or establishments.

\medskip

\begin{figure}[h!]
\includegraphics[scale=.2]{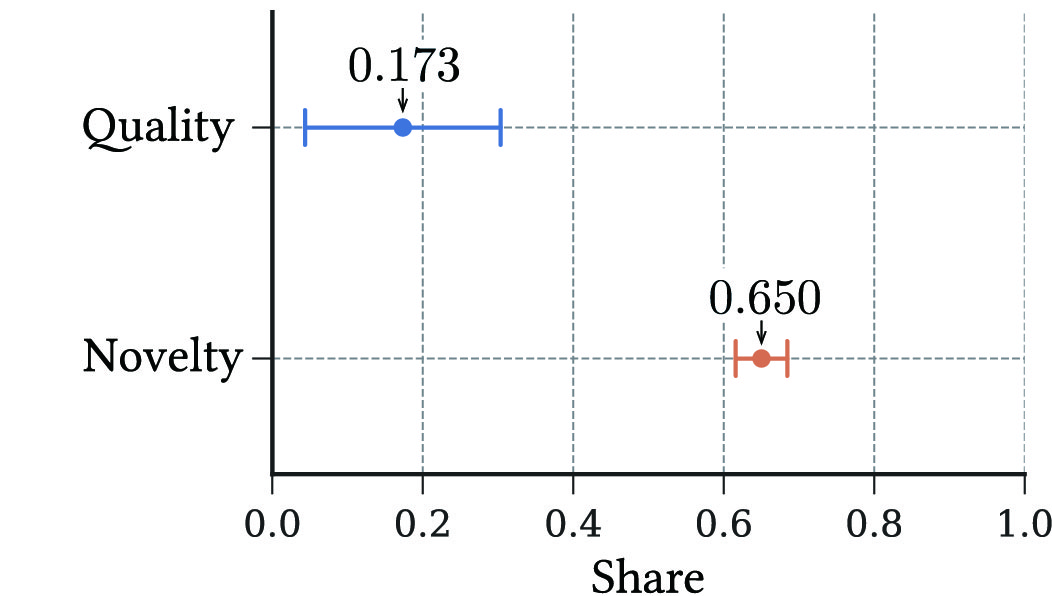}
\caption{Contribution of knowledge exchange on the quality of created knowledge\label{fig:q-q}}
\begin{figurenotes}[\fontsize{10pt}{10pt}\selectfont Notes]
\fontsize{10pt}{10pt}\selectfont
Each horizontal line with the two bounds (indicated by bars) shows the 95\% confidence interval of the share of contribution by collaborators' differentiated knowledge, $\beta^q/\beta$ for $q=\{ \text{quality}, \text{novelty}\}$, where $\beta^q$ and $\beta$ are the coefficients of the differentiated knowledge of collaborators, $\ln k_{it}^D$ in \eqref{eq:quality-equation} and \eqref{eq:baseline}, respectively. The dot in each plot indicates the point estimate of $\beta^q/\beta$. 
The estimation is based on \eqref{eq:baseline} and \eqref{eq:quality-equation} with $\ln k_{it}^D$ instrumented by $\ln k_{it}^{D,{\rm IV}_j}$ for $j = 1,2$ and $3$.
\end{figurenotes}
\end{figure}

\noindent {\it Quantity--quality/novelty decomposition} -- Finally, we turn to the quantity-quality/novelty decomposition of the effect of knowledge exchange on collaborative output based on \eqref{eq:decomposition} and \eqref{eq:quality-equation}.
Figure \ref{fig:q-q} shows the point estimate and 95\% confidence intervals of the estimated shares, $\hat{\beta}^q/\hat{\beta}$, of contributions accruing from knowledge exchange on the average quality and novelty of output.%
\footnote{For each $q=\{ \text{quality}, \text{novelty}\}$, the estimate and confidence interval of $\beta^q/\beta$ are calculated by the generalized method of moments, which simultaneously estimates the baseline model (\ref{eq:baseline}) and the decomposed model (\ref{eq:quality-equation}) with the 2SLS weighting matrix. The comprehensive regression results are relegated to Appendix \ref{app:qq} (Tables \ref{tb:qq-decomposition-citation} and \ref{tb:qq-decomposition-novelty}). The first stage of the regression is shared with \eqref{eq:baseline} and presented in Appendix \ref{app:baseine} (Table \ref{tb:first-stage}).}\ 

We find contrasting roles of knowledge exchange between the quality and novelty of output: $100-17.3 = 83$\%  of its contribution can be attributed to increasing the quantity rather than the quality of research output under the quality-adjusted productivity measure, whereas 65\% of the contribution accrues to increasing the average novelty rather than quantity of research output under the novelty-adjusted measure.
The result indicates that knowledge exchange is comparably more effective for seeking technological novelty than it is for increasing the quality of research output.

Our findings agree with the results of \textcite{Fleming-et-al-ASQ2007} in that cohesiveness of a researcher network has negative effects on the novelty of their developed patents. Since the cohesiveness is expected to be positively correlated with the amount of common knowledge between collaborators, a larger cohesiveness may imply smaller differentiated knowledge within the researcher network, and hence less novelty of output.

\section{Conclusions\label{sec:conclusion}}

\noindent We have shown evidence consistent with the collaborative knowledge creation mechanism proposed by \textcite{Berliant-Fujita-IER2008}. 
To our knowledge, our work is the first to provide micro-econometric evidence for collaborative knowledge creation through the exchange of knowledge at the individual inventor level.
We also found that knowledge exchange tends to raise the novelty comparably more than the quality of collaborative invention.

This evidence has important policy implications. Namely, firms, cities, regions, and countries that promote encounters and collaboration among  inventors across organizations and institutions, despite the possibility of imitations and undesired diffusion, may have better chances to foster innovation through knowledge exchange.

In the future, it will be of interest to further investigate the roles of firms in R\&D. 
As financial resources for R\&D are typically provided by firms, firm-specific patterns of collaborations and R\&D policies could affect the productivity of individual inventors and firms.%
\footnote{See \textcite{Akcigit-Kerr-JPE2018} for an initial attempt in this direction as they distinguish between R\&D that is internal and external to firms and study the firm dynamics that arise from this distinction.}\ 
By matching the addresses of establishments in the patent database with those of the Census of Manufacturers, it is also possible to investigate the impact of patent development on firm productivity.
Second, non-technological diversity of collaborators in terms of, for example, gender, age, and cultural background, may affect productivity. For example, \textcite{Ostergaard-et-al-RP2011} and \textcite{Inui-et-al-DP2014} found a positive influence of gender diversity on innovation of Danish and Japanese firms, respectively.


\clearpage
{
\singlespacing
\printbibliography[title={\textbf{\large REFERENCES}}]
}

\clearpage
\setcounter{page}{1}
\appendix
\begin{center}
\textbf{\large APPENDICES}
\end{center}
\begin{refsection}

\section{\bf \thesection. Data\label{app:data}}
\begin{figure}[h!]
\includegraphics[scale=.2]{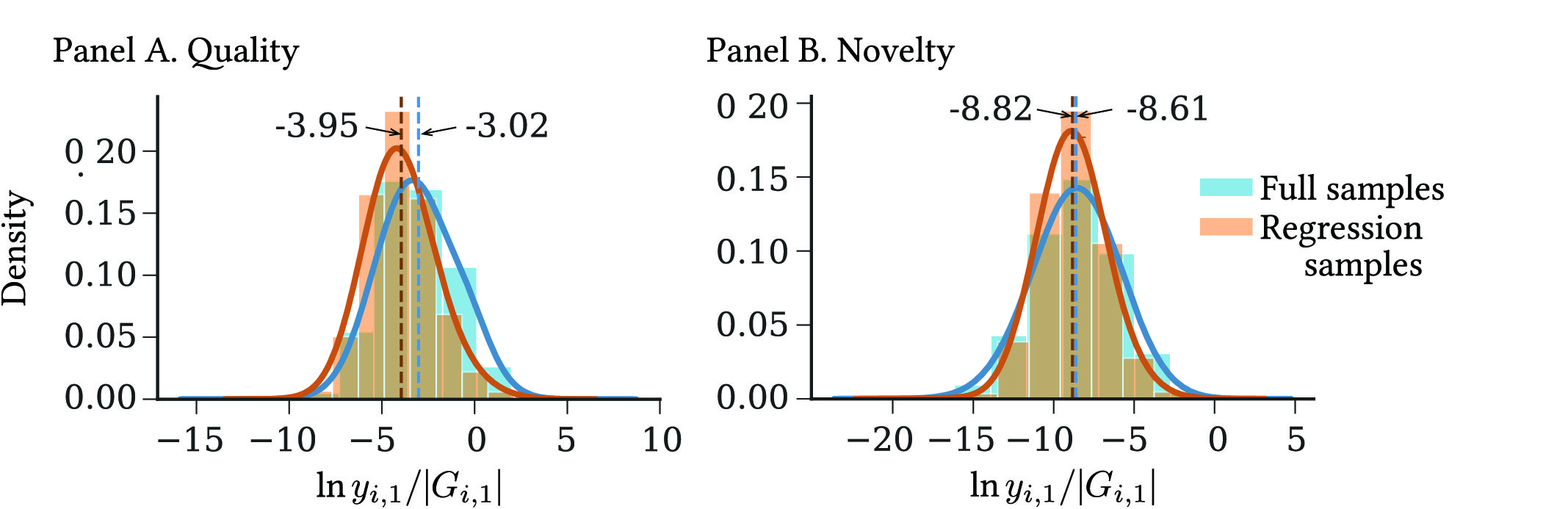}
\caption{Quality and novelty of patents in the regression and full samples (period 1)\label{fig:productivity_dist}}
\begin{figurenotes}[\fontsize{10pt}{10pt}\selectfont Notes]
\fontsize{10pt}{10pt}\selectfont
The figure compares the distributions of the average quality (Panel A) and novelty (Panel B) of patents, $\ln y_{i,1}/|G_{i,1}|$, developed by an inventor through their collaborative knowledge creation between the full set of the 439,260 inventors with at least one collaborator and positive productivity and the set of 29,287 inventors selected for our baseline regressions in Table \ref{tb:bf-model} in period 1. The latter set of inventors appear in both periods, and have at least up to the 5th indirect collaborators in each period.
\end{figurenotes}
\end{figure}

\noindent Tables \ref{tb:descriptive-stat} and \ref{tb:productivity-stat} show the descriptive statistics of the 29,287 focal inventors in $I$ in the panel data used in the regressions, where the latter table collects the variables whose values differ between quality and novelty.
These are the inventors who have at least one collaborator, strictly positive collaborative productivity, and at least 5th-indirect collaborators in each period.
They account for only as 6.7\% and 8.8\% of the full samples of the 439,260 and 331,342 inventors who have at least one collaborator and positive collaborative productivity in periods 1 and 2, respectively.
But, these two sets of inventors do not exhibit any stark difference in the quality and novelty of collaborative output as indicated by Figure \ref{fig:productivity_dist}. 
\begin{table}[tp]
\caption{Descriptive statistics of basic variables\label{tb:descriptive-stat}}
\begin{tabular*}{\textwidth}{lccc}
 \midrule[0.6pt]\midrule[0.6pt]
&&\multicolumn{2}{c@{}}{Period}\\
\cmidrule[0.6pt](rl){3-4}
&&\multicolumn{1}{c}{(1)}&\multicolumn{1}{c}{(2)}\\[3pt]
{\bf Variable}&&\multicolumn{1}{c}{1}&\multicolumn{1}{c}{2}\\
\midrule[0.6pt]
{\bf(1) Number of patents}&$\big |\cup_{i\in I}\mathcal{G}_{it}\big |$&956,711&709,761\\[7pt]
{\bf(2) Number of IPC classes}&$|S_t|$&120&122\\[7pt]
{\bf(3) Number of IPC subclasses}&$|S_t|$&599&588\\[7pt]
{\bf(4) Number of IPC subgroups}&$|S_t|$&31,511&26,424\\[7pt]
{\bf(5) Share of collaborating inventors}&&0.829&0.850\\[3pt]
{\bf (6) Number of collaborators per inventor}&$|N_{it}|$&14.72&12.59\\
&&(11.39)&(9.950)\\[3pt]
{\bf(7) Number of relevant inventors in period $t$}&$|I_t|$&29,287&29,287\\[7pt]
{\bf(8) Number of direct \& indirect collaborators}&$\left|\cup_{\ell = 0}^5\cup_{i\in I_t} N_{it}^\ell \right|$&434,555&283,860\\[7pt]
{\bf(9) Number of inventors per patent}&$| G_{jt} |$&2.968&3.047\\
&&(1.841)&(1.906)\\[3pt]
{\bf (10) Number of patents per inventor}&$|\mathcal{G}_{it}|$&19.23&14.60\\
&&(19.23)&(14.60)\\[3pt]
{\bf (11) Number of IPC sections per inventor}&$|S_t|$&2.223&1.993\\
&&(1.061)&(0.974)\\[3pt]
{\bf (12) Number of IPC classes per inventor}&$|S_t|$&3.318&2.776\\
&&(2.136)&(1.792)\\[3pt]
{\bf (13) Number of IPC subclasses per inventor}&$|S_t|$&4.193&3.452\\
&&(2.959)&(2.496)\\[3pt]
{\bf (14) Number of IPC subgroups per inventor}&$|S_t|$&8.681&6.737\\
&&(6.596)&(5.479)\\[3pt]
\midrule[0.7pt]
\end{tabular*}
\begin{tablenotes}[\fontsize{10pt}{10pt}\selectfont Notes]
\fontsize{10pt}{10pt}\selectfont
This table shows the descriptive statistics of the basic variables of the 29,287 focal inventors used in our regressions. Numbers in parentheses are standard deviations.
\end{tablenotes}
\end{table}

\begin{table}[h!]
\caption{Descriptive statistics of quality and novelty variables\label{tb:productivity-stat}}
\centering
\begin{tabular*}{\textwidth}{p{1in}@{\extracolsep{\fill}}ccccc} \midrule[0.6pt]\midrule[0.6pt]
&& \multicolumn{2}{c@{}}{Panel A. Quality}&\multicolumn{2}{c@{}}{Panel B. Novelty}  \\[2pt]
\cmidrule[0.6pt](lr){3-4}\cmidrule[0.6pt](lr){5-6}
 &  & (1) & (2) & (3) & (4)\\
{\bf Period}&&\multicolumn{1}{c}{1}&\multicolumn{1}{c}{2}&\multicolumn{1}{c}{1}&\multicolumn{1}{c}{2}\\
\midrule[0.6pt]
{\bf Patent value}&$g_{j}$&0.674&0.436&0.009&0.007\\
&&(3.131)&(1.612)&(0.044)&(0.039)\\[3pt]
{\bf Inventer}&$\bar{y}_{it}$&5.134&2.523&0.062&0.038\\
{\bf  productivity}&&(11.50)&(4.568)&(0.147)&(0.101)\\[3pt]
{\bf Pairwise}&$y_{it}$&0.516&0.319&0.006&0.004\\
{\bf  productivity}&&(1.209)&(0.837)&(0.017)&(0.020)\\[3pt]
{\bf Differentiated}&$k_{it}^D$&3.452&1.764&0.040&0.026\\
{\bf  knowledge}&&(5.210)&(2.387)&(0.056)&(0.047)\\[3pt]
\midrule[0.7pt]
\end{tabular*}
\begin{tablenotes}[\fontsize{10pt}{10pt}\selectfont Notes]
\fontsize{10pt}{10pt}\selectfont
This table shows the descriptive statistics of the quality and novelty related variables of the 29,287 focal inventors used in our regressions. Numbers in parentheses are standard deviations.
\end{tablenotes}
\end{table}


\section{\bf \thesection. IPC\label{app:ipc}}
\noindent The IPC classifies technologies into eight sections: A (human necessities), B (performing operations; transporting),$\ldots$, H (electricity). These sections are divided into classes such as A01 (agriculture; forestry; animal husbandry; hunting; trapping; fishing) and then into subclasses such as A01C (planting; sowing; fertilizing). Each subclass is further divided into groups, e.g., A01C1 (apparatus, or methods of use thereof, for testing or treating seed, roots, or the like, prior to sowing or planting), and then into subgroups, e.g., A01C1/06 (coating or dressing seed) and A01C1/08 (immunizing seed). 
The IPC's labeling scheme is consistent over time, and a newly introduced category is basically associated with a new technology (e.g., the classes B81 for microtechnology and B82 for nanotechnology introduced in 2000). 
As another example, the shale revolution in the late 2000s in the United States was made possible by some key innovations in excavation technology that mainly belong to a new subclass C09K (compositions for drilling of boreholes or wells; compositions for treating boreholes or wells) that was split from E21B (earth or rock drilling; obtaining oil, gas, water, soluble or meltable materials or a slurry of minerals from wells) in 2006.
If there are no fundamental changes in technology in a given category, the classification remains the same (e.g., A47C for furniture; domestic articles or appliances; coffee mills; spice mills; suction cleaners in general). 
Taken together, the set of technological categories specified in the IPC at a given point in time roughly represents the set of the state-of-the-art technologies at that time, and hence makes an appropriate proxy for the set of technological knowledge.

We have 121, 609, and 40,691 (123, 616, and 38,339) relevant IPC classes, subclasses, and subgroups, respectively for period 1 (period 2), associated with the applied patents in our data. 


\section{\bf \thesection. Geographic neighborhood factors\label{app:locational-factors}}
\noindent This section describes UAs and gives precise definitions for the measures of geographic neighborhood factors discussed in Section \ref{sec:data}. The descriptive statistics of the neighborhood factors are shown in Table \ref{tb:locational-factors}.

\begin{figure}[h!]%
\begin{center}
\includegraphics[scale=.12]{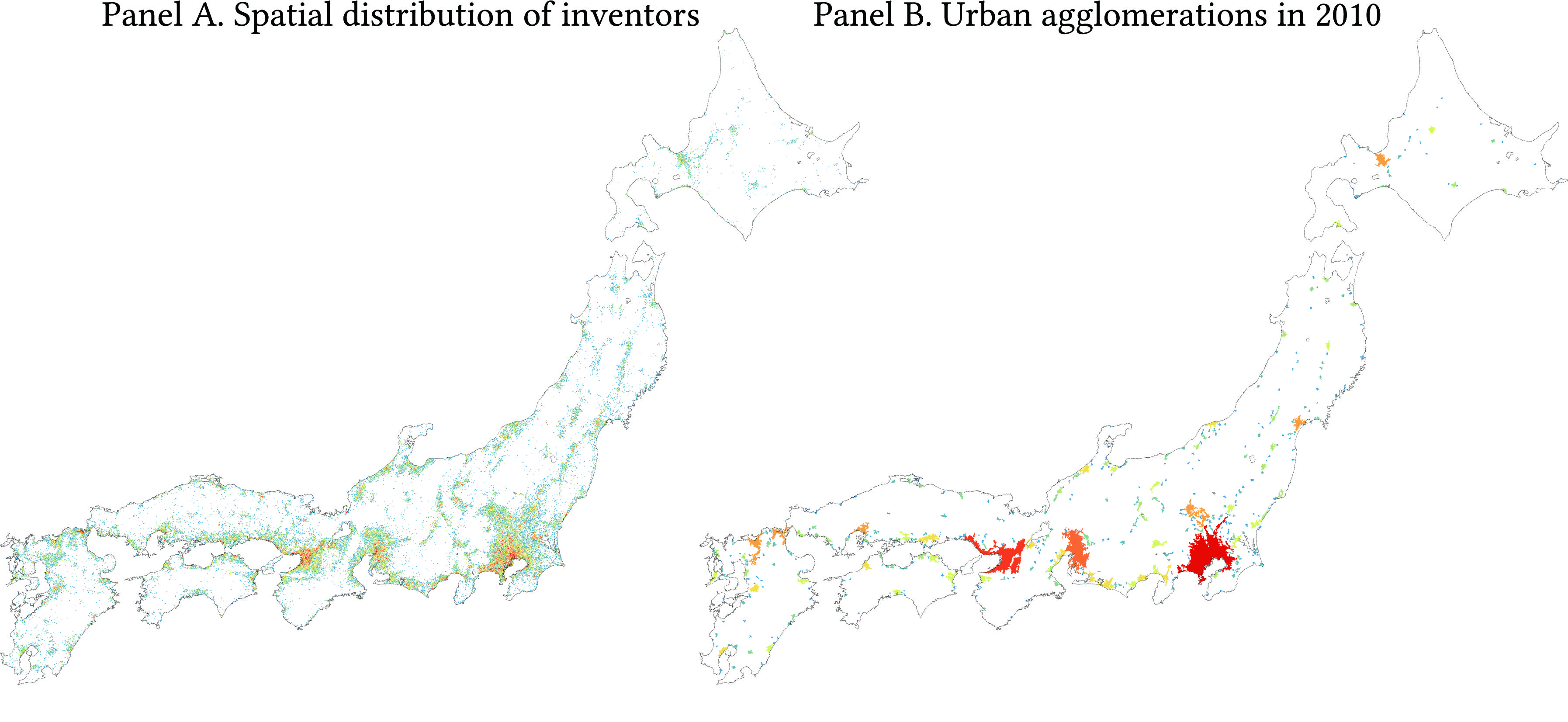}
\bigskip
\caption{Spatial distribution of inventors and UAs\label{fig:ua}}%
\end{center}
\begin{figurenotes}[\fontsize{10pt}{10pt}\selectfont Notes]
\fontsize{10pt}{10pt}\selectfont The map of Japan in the figure includes four major islands (Honshu, Kyushu, Shikoku and Hokkaido) together with other islands connected to at least one of these major islands by road. Panel A shows the geographic distribution of inventors in Japan, and Panel B shows the distribution of population over 453 UAs indicated by colored patches on the map. In both panels, warmer colors indicate higher population density.
The population data at the 1 km-by-1 km grid level are obtained from the Population Census (\citeyear{Pop-Census}) by  MIAC.
\end{figurenotes}
\end{figure}

\smallskip
\noindent{\it UAs} -- Panels A and B in Figure \ref{fig:ua} show the spatial distribution of inventors and 453 UAs as of 2010, respectively, where the warmer colors in each panel indicate higher population density.
Each inventor is assigned to the closest UA if there is any UA within 10 km of their location.

\smallskip
\noindent{\it Inventor concentration} -- The local population, $a_{it}^{\text{INV}}$, of inventors within a given distance, $\bar{d}$, of the location of inventor $i$ is defined as
 \begin{equation}
 	a_{it}^{\text{INV}} = \left|\big\{j\in I_t\backslash N_{it}\:\::\:\: d(i,j) < \bar{d} \big\}\right|,
 \end{equation}
where $d(i,j)$ represents the great-circle distance between inventors $i$ and $j$ (rows 1--3, Table \ref{tb:locational-factors}). To evaluate the pure spillover effects, this population excludes the collaborators, $N_{it}$, of $i$.%
\footnote{The effects of externalities from the nearby inventors and firms that have been recognized in the literature (e.g., \citeauthor{Jaffe-Trajtenberg-Henderson-QJE1993}, \citeyear{Jaffe-Trajtenberg-Henderson-QJE1993};
\citeauthor{Thompson-Fox-Kean-2005}, \citeyear{Thompson-Fox-Kean-2005}; \citeauthor{Murata-et-al-RES2014}, \citeyear{Murata-et-al-RES2014}; \citeauthor{Kerr-Kominers-REStat2015}, \citeyear{Kerr-Kominers-REStat2015}).}\ %
%
%

\smallskip
\noindent {\it R\&D expenditure} -- Focusing on manufacturing, we first aggregate firm-level R\&D expenditure at the industry level according to the three-digit Japanese Standard Industry Classification (SIC) in each period $t$. Denote the industry-level R\&D expenditure (in million yen) by $v_m$ for each industry $m\in M_t$, where $M_t$ is the set of three-digit manufacturing industries in period $t$.%
\footnote{Data on R\&D expenditure at the firm level are available for firms with at least four employees for every year from 2000 to 2009 from the Survey of Research and Development.}\ 

Next, from the micro data of the Establishment and Enterprise Census as well as the Economic Census (MIAC, \citeyear{Establishment-Census,Economic-Census}), we find the set of establishments, $E_{mt}$, in each industry $m\in M_t$ in period $t$, and compute the employment share, $e_{kt}$, of each establishment $k\in E_{mt}$ within industry $m$. 

Assuming that the R\&D expenditure of each establishment in each industry is proportional to the employment size of the establishment, the value of R\&D expenditure of each establishment in period $t$ is approximated by $v_{mt}e_{mt}$. Assuming that the R\&D expenditure in the previous period $t-1$ affects the productivity of inventors in the current period $t$, the R\&D around inventor $i$ in period $t$ is given as follows (rows 4--6, Table \ref{tb:locational-factors}):%
	\footnote{The R\&D expenditure values are obtained from the Survey of Research and Development (\citeyear{RDSurvey}) by MIAC and from METI Basic Survey of Japanese Business Structure and Activities (\citeyear{Kikatsu}) by METI.}\ 
\begin{equation}
	a_{it}^{\text{R\&D}} = \sum_{m\in M_{t}}\hspace{5pt}\sum_{k\in \{ j\in E_m \::\: d(i,j) < \bar{d} \}} \:\: v_{m,t-1}e_{k,t-1}.
\end{equation}
$a_{it}^{\rm{R\&D}}$ is naturally expected to influence patent development.

\smallskip
\noindent{\it Manufacturing concentration} -- Assuming that the employment size and output of an establishment correlate with demand for new knowledge, we proxy the local market size for an invented technology around inventor $i$ by the local manufacturing employment and output around $i$:%
\footnote{Another interpretation of $a_{it}^\text{MNF}$ is the spillover from manufacturing concentration around $i$ in period $t$.}
\begin{equation}
	a_{it}^{\text{MNF}_j} = \sum_{k\in \{ j\in E_t \::\: d(i,j) < \bar{d} \}} e_{kt}\,
\end{equation}
where $E_t= \cup_{m\in M_t} E_{mt}$, and $e_{kt}$ represents the total output value (employment) of establishment $k$ for $j = o$ ($j = e$) (rows 7--12, Table \ref{tb:locational-factors}).%
\footnote{The manufacturing employment values are obtained from the Establishment and Enterprise Census for (\citeyear{Establishment-Census}) and Economic Census for Business Frame  (\citeyear{Economic-Census}) by MIAC; the manufacturing output values are obtained from the micro data of the Census of Manufacturers (\citeyear{Manu-Census}) and Economic Census for Business Frame (\citeyear{Economic-Census}) by MIAC.}

\smallskip
\noindent{\it Residential population}  -- The local residential population is defined as
\begin{equation}
	a_{it}^{\text{POP}} = \sum_{k\in \{ j\in R\::\: d(i,j) < \bar{d} \}} r_{kt}\,
\end{equation}
where $R$ represents the set of 1km-by-1km cells covering the relevant location space in Japan; the centroid of each cell is considered to be the representative location of the cell in measuring the distance from the cell; $r_{kt}$ is the residential population in cell $k\in R$ at the beginning of period $t$ (rows 13--15, Table \ref{tb:locational-factors}).%
\footnote{The residential population in the 1 km-by-1 km cells is available from the Population Census (\citeyear{Pop-Census}) by MIAC.}\

\begin{center}
\begin{table}[tp]
\caption{Descriptive statistics of the geographic neighborhood factors\label{tb:locational-factors}}
\begin{tabular*}{\textwidth}{p{2in}@{\extracolsep{\fill}}ccc} 
\midrule[0.6pt]\midrule[0.6pt]
&&\multicolumn{1}{c}{\bf (1)}&\multicolumn{1}{c}{\bf (2)}\\
 {\bf Period}&&\multicolumn{1}{c}{1}  & \multicolumn{1}{c}{2}  \\
\midrule[0.6pt]
{\bf Inventor population} &1km&   5,750 &  5,629\\
  &&(7,225)& (7,282)\\[3pt]
    &5km & 31,026& 30,158\\
  &&(42,143)&(42,269)  \\[3pt]
  &10km & 70,720&66,011\\
  &&(79,277)&(77,330)  \\[3pt]
  {\bf R\&D investment } &1km& 10,454&  18,480\\
  &&(78,020) & (180,284) \\ [3pt]
    &5km&150,581 &278,911\\
  &&(338,668)&(703,381)\\[3pt]  
    &10km&300,256 &520,066\\
  &&(466,130)&(920,505)\\[3pt]
 {\bf Manufacturing employment} &1km& 2,240 &6,676\\
&&(1,505)&(7,106)\\[3pt]
 &5km&52,974 &76,491\\
&&(32,395)&(74,655)\\[3pt]
 &10km&182,597 &212,371\\
&&(106,414)&(166,473)\\[3pt]
 {\bf Manufacturing output}&1km&  21,801,942 &20,774,589\\
\hspace{25pt} (in thousand)    && (58,182,730) &(83,883,736) \\[3pt]
  &5km&158,183,183 &104,957,604\\
&&(129,167,825)&(129,388,708)\\[3pt]
 &10km&445,908,195 &317,846,559\\
&&(255,976,915)&(226,259,080)\\[3pt]
{\bf Residential population}&5km &  595,461& 615,722\\
&&(386,442)&(399,930)\\[3pt]
&10km&2,100,541  &2,156,271\\
&&(1,388,078)&(1,432,171)\\[3pt]
&20km& 6,386,959 & 6,573,357\\
&&(4,252,098)&(4,416,168)\\[3pt]
\midrule[0.7pt]
\end{tabular*}
\begin{tablenotes}[\fontsize{10pt}{10pt}\selectfont Notes]
\fontsize{10pt}{10pt}\selectfont 
Numbers are the average values with standard deviations in parentheses.
\end{tablenotes}
\end{table}
\end{center}

\clearpage
\section{\bf \thesection. Baseline Regression Results\label{app:baseine}}

\noindent Table \ref{tb:first-stage} summarizes the first-stage results for the 2SLS-IV regressions for model \eqref{eq:baseline}. 
Tables \ref{tb:bf-model-citation-full-result} and \ref{tb:bf-model-novelty-full-result} add the second-stage results for the estimated effects of geographic neighborhood factors to Table \ref{tb:bf-model} under quality- and novelty-adjusted productivities, respectively.

For geographic neighborhood factors, we generally have expected results. 
In particular, the nearby concentrations of R\&D expenditure and manufacturing employment have persistent positive results for all specifications. 
The fact that the nearby concentration of inventors have only weak or even negative effects suggests that the knowledge exchange is the primary cause of agglomeration for R\&D activities.
The effects of manufacturing output do not seem to have clear additional factors to manufacturing employment.
The negative effects of residential population indicate that the density of R\&D expenditure and manufacturing matters.

\begin{table}[tp]
\centering
\caption{First stage regression results for the Berliant-Fujita model\label{tb:first-stage}}
\begin{tabular*}{\textwidth}{p{1in}@{\extracolsep{\fill}}cccc} 
\midrule[0.6pt]\midrule[0.6pt]
& \multicolumn{4}{c@{}}{Panel A. Quality} \\[2pt]
\cmidrule[0.6pt]{2-5}
Variable & (1) IV3-5 & (2) IV3 & (3) IV4 & (4) IV5\\
\midrule[0.6pt]
$\ln k_{it}^{D,\rm{IV}_3}$ & 0.373 & 0.364 &  &  \\
 & (0.0466) & (0.0273) &  &  \\
$\ln k_{it}^{D,\rm{IV}_4}$ & 0.0384 &  & 0.257 &  \\
 & (0.0464) &  & (0.0398) &  \\
$\ln k_{it}^{D,\rm{IV}_5}$ & -0.0896 &  &  & 0.139 \\
 & (0.0261) &  &  & (0.0482) \\
$\ln k_{it}$ & 0.127 & 0.127 & 0.131 & 0.134 \\
 & (0.0481) & (0.0473) & (0.0542) & (0.0595) \\
$(\ln k_{it})^2$ & -0.0402 & -0.0403 & -0.0423 & -0.0440 \\
 & (0.0156) & (0.0155) & (0.0185) & (0.0198) \\
\midrule[0.6pt]
$R^2$ & 0.0336 & 0.0328 & 0.0130 & 0.00311 \\
$F^{\mbox{eff}}$ & 52.96 & 179 & 41.90 & 8.413 \\
\midrule[0.6pt]
\\
& \multicolumn{4}{c@{}}{Panel B. Novelty} \\[2pt]
\cmidrule[0.6pt]{2-5}
Variable & (1) IV3-5 & (2) IV3 & (3) IV4 & (4) IV5\\
\midrule[0.6pt]
$\ln k_{it}^{D,\rm{IV}_3}$ & 0.443 & 0.466 &  &  \\
 & (0.0204) & (0.0147) &  &  \\
$\ln k_{it}^{D,\rm{IV}_4}$& 0.0396 &  & 0.397 &  \\
 & (0.0168) &  & (0.0148) &  \\
$\ln k_{it}^{D,\rm{IV}_5}$ & -0.00724 &  &  & 0.319 \\
 & (0.0332) &  &  & (0.0248) \\
$\ln k_{it}$ & 0.146 & 0.146 & 0.161 & 0.165 \\
 & (0.0152) & (0.0153) & (0.0174) & (0.0207) \\
$(\ln k_{it})^2$ & -0.0741 & -0.0739 & -0.0782 & -0.0793 \\
 & (0.00854) & (0.00858) & (0.0104) & (0.0124) \\
\midrule[0.6pt]
$R^2$ & 0.0546 & 0.0545 & 0.0311 & 0.0164 \\
$F^{\mbox{eff}}$ &  332.9 & 1009 & 720.8 & 166.3 \\
\midrule[0.7pt]
\end{tabular*}
\begin{tablenotes}[\fontsize{10pt}{10pt}\selectfont Notes]
\fontsize{10pt}{10pt}\selectfont 
The table shows the first-stage results for the key variables in the 2SLS-IV regressions for model \eqref{eq:baseline}. 
(i) The number of observations is 58,574 (29,287 inventors$\times$2 time periods) in all the regressions. 
(ii) The dependent variable is log of the differentiated knowledge of collaborators of inventor $i$, $\ln k_{it}^D$. It is defined in terms of cited counts and novelty of patents in Panels A and B, respectively. 
(iii) Explanatory variables shown are the average differentiated knowledge of indirect collaborators, $\ln k_{it}^{D,\rm{IV}_\ell}$ for $\ell = 3,4$ and $5$; the first- and second-order effects of the research scope, $\ln k_{it}$, of an inventor.
(iv) In all the regressions, year and IPC class fixed effects, as well as a variety of neighborhood effects are controlled. Neighborhood effects include the sizes of concentration of inventors, R\&D expenditure, manufacturing employment and output within a circle of 1-km radius, and population within a circle of 20-km radius around a given inventor.
(v) The last row in each panel reports \cite{Olea-Pflueger-JBES2013}'s effective first-stage $F$-statistic.
\end{tablenotes}
\end{table}
\begin{table}[tp]
\centering
\caption{Regression results for the BF model under quality-based productivity\label{tb:bf-model-citation-full-result}}
\begin{tabular*}{\textwidth}{p{1in}@{\extracolsep{\fill}}ccccc} 
\midrule[0.6pt]\midrule[0.6pt]\\[-5pt]
Variable &(1) OLS & (2) IV3-5 & (3) IV3 & (4) IV4 & (5) IV5\\[2pt]
\midrule[0.6pt]
$\ln k_{it}^D$ & 0.166 & 0.334 & 0.340 & 0.392 & 0.488 \\[-2pt]
& (0.0111) & (0.0444) & (0.0440) & (0.0565) & (0.118) \\[2.5pt]
$\ln k_{it}$ & 0.163 & 0.140 & 0.139 & 0.132 & 0.119 \\[-2pt]
& (0.0326) & (0.0213) & (0.0212) & (0.0199) & (0.0192) \\[2.5pt]
$(\ln k_{it})^2$ & -0.0744 & -0.0669 & -0.0666 & -0.0643 & -0.0600 \\[-2pt]
 & (0.0116) & (0.00818) & (0.00816) & (0.00765) & (0.00721) \\
 $\ln a_{it}^{\text{INV}}$ & -0.0668 & -0.108 & -0.109 & -0.122 & -0.146 \\[-2pt]
 & (0.0871) & (0.0995) & (0.0997) & (0.0977) & (0.108) \\
 $\ln a_{it}^{\text{R\&D}}$ & 0.0176 & 0.0124 & 0.0122 & 0.0106 & 0.00763 \\[-2pt]
 & (0.00711) & (0.00639) & (0.00631) & (0.00564) & (0.00503) \\
 $\ln a_{it}^{\text{MNF}_e}$  & 0.0906 & 0.0852 & 0.0850 & 0.0833 & 0.0803 \\[-2pt]
 & (0.0119) & (0.0116) & (0.0116) & (0.0111) & (0.0118) \\
 $\ln a_{it}^{\text{MNF}_o}$ & 0.0206 & 0.0154 & 0.0153 & 0.0136 & 0.0107 \\[-2pt]
 & (0.0139) & (0.0173) & (0.0173) & (0.0175) & (0.0195) \\
 $\ln a_{it}^{\text{POP}}$ & -3.792 & -3.697 & -3.694 & -3.664 & -3.610 \\[-2pt]
 & (1.055) & (0.832) & (0.824) & (0.761) & (0.670) \\
 $\tau_1$ & 0.218 & 0.104 & 0.0996 & 0.0643 & -0.000898 \\[-2pt]
& (0.0485) & (0.0313) & (0.0314) & (0.0511) & (0.0799) \\
 \cmidrule[0.6pt]{1-6}
$R^2$ & 0.123 & 0.106 & 0.104 & 0.091 & 0.059 \\
$F^{\mbox{eff}}$  & & 52.96 & 179 & 41.90 & 8.413 \\
\multicolumn{2}{l@{}}{Critical $F^{\mbox{eff}}$-value}  & 20.09 & 23.11 & 23.11 & 23.11 \\ 
\multicolumn{2}{l@{}}{Hansen $J$ $p$-value}   & 0.681 &  &  &  \\
\midrule[0.7pt]
\end{tabular*}
\begin{tablenotes}[\fontsize{10pt}{10pt}\selectfont Notes]
\fontsize{10pt}{10pt}\selectfont
(i) The number of observations is 58,574 (29,287 inventors$\times$2 time periods) in all the regressions. The inventors included have up to at least the 5th indirect collaborators in order to construct instruments for an endogenous variable. 
(ii) The dependent variable is log of the average pairwise productivity of an inventor, $\ln y_{it}$. It is defined in terms of cited counts and novelty of patents in Panels A and B, respectively. 
(iii) Explanatory variables shown are the average differentiated knowledge of collaborators, $\ln k_{it}^D$; the first- and second-order effects of the research scope, $\ln k_{it}$, of an inventor.
(iv) Column 1 shows the result from the OLS regression, while columns 2-5 show the results of IV regressions, where $\ln k_{it}^D$ is treated as an endogenous variable, and is instrumented by the same variable of the 3rd-5th, 3rd, 4th and 5th indirect collaborators of inventor $i$ in columns 2, 3, 4 and 5, respectively. 
(v) In all the regressions, inventor and IPC class fixed effects are controlled. Neighborhood effects include the sizes of concentration of inventors, R\&D expenditure, manufacturing employment and output within a circle of 1-km radius, and population within a circle of 20-km radius around a given inventor. 
(vi) The third to last row reports \cite{Olea-Pflueger-JBES2013}'s effective first-stage $F$-statistic.
(vii) The second to last row indicates 5\% critical value of the effective $F$-statistic with a Nagar bias threshold $\tau=10\%$.
(viii) The last row in each panel indicates the $p$-value of \cite{Hansen-ECTA1982}'s $J$ test.
(ix) Robust standard errors in parentheses are clustered on urban agglomeration.
\end{tablenotes}
\end{table} 
\begin{table}[h!]
\centering
\caption{Regression results for the BF model under novelty-based productivity \label{tb:bf-model-novelty-full-result}}
\begin{tabular*}{\textwidth}{p{1in}@{\extracolsep{\fill}}ccccc} 
\midrule[0.6pt]\midrule[0.6pt]\\[-5pt]
Variable & (1) OLS & (2) IV3-5 & (3) IV3 & (4) IV4 & (5) IV5\\[2pt]
\midrule[0.6pt]
$\ln k_{it}^D$ & 0.217 & 0.480 & 0.478 & 0.511 & 0.495 \\[-2pt]
& (0.00693) & (0.0403) & (0.0409) & (0.0335) & (0.0491) \\
$\ln k_{it}$& 0.235 & 0.189 & 0.190 & 0.184 & 0.187 \\[-2pt]
& (0.0187) & (0.0166) & (0.0166) & (0.0174) & (0.0202) \\
$(\ln k_{it})^2$ & -0.182 & -0.161 & -0.161 & -0.158 & -0.160 \\[-2pt]
& (0.0148) & (0.00924) & (0.00920) & (0.00982) & (0.00916) \\
 $\ln a_{it}^{\text{INV}}$ & 0.262 & 0.112 & 0.113 & 0.0941 & 0.104 \\[-2pt]
  & (0.121) & (0.119) & (0.120) & (0.112) & (0.109) \\
  $\ln a_{it}^{\text{R\&D}}$ & 0.0442 & 0.0307 & 0.0308 & 0.0291 & 0.0300 \\[-2pt]
  & (0.0192) & (0.0158) & (0.0158) & (0.0153) & (0.0146) \\
  $\ln a_{it}^{\text{MNF}_e}$ & 0.00718 & 0.0326 & 0.0324 & 0.0356 & 0.0340 \\[-2pt]
  & (0.0155) & (0.0155) & (0.0156) & (0.0153) & (0.0171) \\
  $\ln a_{it}^{\text{MNF}_o}$ & -0.0123 & -0.0114 & -0.0115 & -0.0113 & -0.0114 \\[-2pt]
  & (0.00637) & (0.0100) & (0.00998) & (0.0107) & (0.0103) \\
 $\ln a_{it}^{\text{POP}}$ & -0.145 & -1.299 & -1.291 & -1.436 & -1.363 \\[-2pt]
  & (0.980) & (0.948) & (0.950) & (0.896) & (0.897) \\
$\tau_1$ & 0.196 & 0.0194 & 0.0206 & -0.00162 & 0.00946 \\[-2pt]
 & (0.0450) & (0.0558) & (0.0562) & (0.0497) & (0.0592) \\
 \cmidrule[0.6pt]{1-6}
$R^2$& 0.175 & 0.134 & 0.135 & 0.124 & 0.129 \\
$F^{\mbox{eff}}$  && 332.9 & 1009 & 720.8 &  166.3\\
\multicolumn{2}{l@{}}{Critical $F^{\mbox{eff}}$-value} & 17.85 & 23.11 & 23.11 & 23.11 \\
\multicolumn{2}{l@{}}{Hansen $J$ $p$-value} & 0.113 &  &  &  \\
\midrule[0.7pt]
\end{tabular*}
\begin{tablenotes}[\fontsize{10pt}{10pt}\selectfont Notes]
\fontsize{10pt}{10pt}\selectfont
(i) The number of observations is 58,574 (29,287 inventors$\times$2 time periods) for all regressions. The inventors included have up to at least the 5th indirect collaborators in order to construct instruments for an endogenous variable. 
(ii) The dependent variable is log of the average pairwise productivity of an inventor, $\ln y_{it}$. It is defined in terms of cited counts and novelty of patents in Panels A and B, respectively. 
(iii) The explanatory variables shown are the average differentiated knowledge of collaborators, $\ln k_{it}^D$; the first- and second-order effects of the research scope, $\ln k_{it}$, of an inventor.
(iv) Column 1 shows the result from the OLS regression, while columns 2-5 show the results of IV regressions, where $\ln k_{it}^D$ is treated as an endogenous variable, and is instrumented by the same variable of the 3rd-5th, 3rd, 4th and 5th indirect collaborators of inventor $i$ in columns 2, 3, 4 and 5, respectively. 
(v) In all the regressions, inventor and IPC class fixed effects are controlled. Neighborhood effects include the sizes of concentration of inventors, R\&D expenditure, manufacturing employment and output within a 1-km radius, and population within a 20-km radius around a given inventor. 
(vi) The third to last row reports \cite{Olea-Pflueger-JBES2013}'s effective first-stage $F$-statistic.
(vii) The second to last row indicates 5\% critical value of the effective $F$-statistic with a Nagar bias threshold $\tau=10\%$.
(viii) The last row in each panel indicates the $p$-value of \cite{Hansen-ECTA1982}'s $J$ test.
(ix) Robust standard errors in parentheses are clustered on urban agglomeration.
\end{tablenotes}
\end{table} 


\clearpage
\section{\bf \thesection. Robustness\label{app:robustness}}

\noindent 
The estimated coefficient of $\ln k_{it}^D$ in \eqref{eq:baseline} might reflect not only the effects of differentiated knowledge of collaborators, but also those of  time-varying factors specific to the inventors' own and collaborators' firms or establishments. Such factors include R\&D environment and productivity externality (peer effects) from (possibly non-collaborating) inventors that vary across firms or establishments, and their research affiliations.
This section investigates how much of the estimated coefficient of $\ln k_{it}^D$ in \eqref{eq:baseline} is explained by the common factors at the level of a firm, an establishment or their research affiliation, rather than by the direct exchange of knowledge at the inventor level.


\smallskip

\noindent {\it Firm/establishment size and scope} -- We first consider factors that correlate with the size and research scope of the firm to which an inventor belongs.
Let $F_{it}$ be the set of inventors who belong to the same firm as inventor $i$ at some point in period $t$, and let $F_{-i,t} \equiv F_{it} \backslash \left(N_{it}\cup \{i\}\right)$, that is $F_{it}$  excluding $i$ and their collaborators.
The \emph{firm size}, $f_{it} = |F_{-i,t}|$ captures the magnitude of R\&D activities within the firm to which inventor $i$ belongs; however, outside the projects, $i$ and their collaborators are directly involved.
Given that more than 80\% of collaboration occurs within a firm on average, the variation in $k_{it}^D$ may simply reflect firm size in period $t$.
We also include the \emph{research scope} of the firm to which inventor $i$ belongs defined by $s_{it}^f = |\cup_{j\in F_{it}} S_{jt}\backslash (\cup_{u\in N_{it}\cup \{i\}}S_{ut})|$.
It counts the number of distinct technological categories associated with the patents developed in the inventor $i$'s firm, while excluding those associated with the patents developed by $i$ and their collaborators.
The values of $f_{it}$ and $s_{it}^f$ reflect the potential scale effect of a firm; for example, the availability of common research facilities, funding, and other sources of increasing returns as well as spillover.
In a similar manner, we can define the set $E_{it}$ of inventors who belong to the same establishment as inventor $i$ in period $t$ and set the \emph{establishment size} $e_{it} = |E_{-it}|$ as well as the research scope $s^e_{it}  = |\cup_{j\in E_{it}} S_{jt}\backslash (\cup_{u\in N_{it}\cup \{i\}}S_{ut})|$ of their establishment.
\begin{table}[tp]
\centering
\caption{First-stage regression results with firm/establishment size and research scope effects\label{tb:first-firm}}
\begin{tabular*}{\textwidth}{p{1in}@{\extracolsep{\fill}}cccc} \midrule[0.6pt]\midrule[0.6pt]
& \multicolumn{2}{c@{}}{Panel A. Quality}&\multicolumn{2}{c@{}}{Panel B. Novelty}  \\[2pt]
\cmidrule[0.6pt](lr){2-3}\cmidrule[0.6pt](lr){4-5}
Variable & (1) & (2) & (1) & (2)\\
\midrule[0.6pt]
$\ln k_{it}^{D,\rm{IV}_3}$ & 0.346 & 0.288 & 0.397 & 0.313 \\
 & (0.0179) & (0.0177) & (0.0196) & (0.0197) \\
$\ln k_{it}^{D,\rm{IV}_4}$ & 0.0359 & 0.0434 & 0.0134 & 0.0212 \\
 & (0.0238) & (0.0235) & (0.0260) & (0.0256) \\
$\ln k_{it}^{D,\rm{IV}_5}$ & -0.0933 & -0.0795 & -0.0236 & -0.0172 \\
 & (0.0215) & (0.0214) & (0.0234) & (0.0232) \\
 $\ln k_{it}$ & 0.105 & 0.0833 & 0.116 & 0.0890 \\
 & (0.0180) & (0.0178) & (0.0216) & (0.0213) \\
$(\ln k_{it})^2$ & -0.0362 & -0.0261 & -0.0665 & -0.0530 \\
 & (0.00549) & (0.00540) & (0.00664) & (0.00651) \\
 $\ln f_{it}$ & -0.530 &  & -0.704 &  \\
 & (0.0476) &  & (0.0602) &  \\
$\ln s_{it}^f$ & 0.825 &  & 1.365 &  \\
 & (0.0536) &  & (0.0690) &  \\
$\ln e_{it}$ &  & -0.511 &  & -0.542 \\
 &  & (0.0465) &  & (0.0555) \\
$\ln s_{it}^e$ &  & 0.985 &  & 1.375 \\
 &  & (0.0403) &  & (0.0500) \\
\midrule[0.6pt]
$R^2$ & 0.0278 & 0.0205 & 0.0377 & 0.0253 \\
$F^{\mbox{eff}}$ & 512.1 & 512.1 & 823.1 & 823.1\\
\midrule[0.7pt]
\end{tabular*}
\begin{tablenotes}[\fontsize{10pt}{10pt}\selectfont Notes]
\fontsize{10pt}{10pt}\selectfont 
The table shows the first-stage results for the key variables in the 2SLS-IV regressions for model \eqref{eq:baseline} with firm/establishment size and scope controls.
(i) The number of observations is 58,574 (29,287 inventors$\times$2 time periods) for all regressions. 
(ii) The dependent variable is log of the differentiated knowledge of collaborators of inventor $i$, $\ln k_{it}^D$. It is defined in terms of cited counts and novelty of patents in Panels A and B, respectively. 
(iii) The explanatory variables shown are the average differentiated knowledge of indirect collaborators, $\ln k_{it}^{D,\rm{IV}_\ell}$ for $\ell = 3,4$ and $5$; the first- and second-order effects of the research scope, $\ln k_{it}$, of an inventor; sizes, $f_{it}$ and $e_{it}$, and research scopes (in terms of IPC subgroup), $s_{it}^f$ and $s_{it}^e$, of the firm and establishment to which $i$ belongs.
(iv) In all the regressions, inventor, year and IPC class fixed effects, as well as a variety of neighborhood effects are controlled. Neighborhood effects include the sizes of concentration of inventors, R\&D expenditure, manufacturing employment and output within a 1-km radius, and population within a 20-km radius around a given inventor.
(v) The last row reports \cite{Olea-Pflueger-JBES2013}'s effective first-stage $F$-statistic.
\end{tablenotes}
\end{table} 

\begin{table}[tp]
\centering
\caption{Regression results with firm/establishment size and research scope effects\label{tb:firm-factors}}
\begin{tabular*}{\textwidth}{p{1in}@{\extracolsep{\fill}}cccc} \midrule[0.6pt]\midrule[0.6pt]
& \multicolumn{2}{c@{}}{Panel A. Quality}&\multicolumn{2}{c@{}}{Panel B. Novelty}  \\[2pt]
\cmidrule[0.6pt](lr){2-3}\cmidrule[0.6pt](lr){4-5}
Variable & (1) & (2) & (1) & (2)\\
\midrule[0.6pt]
$\ln k_{it}^D$ & 0.331 & 0.368 & 0.442 & 0.409 \\
 & (0.0455) & (0.0542) & (0.0396) & (0.0495) \\
 $\ln k_{it}$ & 0.142 & 0.145 & 0.187 & 0.186 \\
 & (0.0223) & (0.0223) & (0.0267) & (0.0264) \\
$(\ln k_{it})^2$ & -0.0689 & -0.0685 & -0.162 & -0.161 \\
 & (0.00684) & (0.00683) & (0.00847) & (0.00841) \\
$\ln f_{it}$ & -0.283 &  & -0.373 &  \\
 & (0.0676) &  & (0.0794) &  \\
$\ln s_{it}^f$ & 0.161 &  & 0.445 &  \\
 & (0.0828) &  & (0.105) &  \\
$\ln e_{it}$ &  & -0.250 &  & -0.280 \\
 &  & (0.0640) &  & (0.0712) \\
$\ln s_{it}^e$ &  & 0.0367 &  & 0.420 \\
 &  & (0.0782) &  & (0.0976) \\
\midrule[0.6pt]
$R^2$ & 0.107 & 0.099 & 0.144 & 0.154 \\
$F^{\mbox{eff}}$ & 512.1 & 512.1 & 823.1 & 823.1 \\
\multicolumn{1}{l@{}}{Critical $F^{\mbox{eff}}$-value} & 9.361 & 9.361 & 9.453 & 9.453\\
\multicolumn{1}{l@{}}{Hansen $J$ $p$-value} & 0.424 & 0.433 & 0.757 & 0.711\\
\midrule[0.7pt]
\end{tabular*}
\begin{tablenotes}[\fontsize{10pt}{10pt}\selectfont Notes]
\fontsize{10pt}{10pt}\selectfont
(i) The number of observations is 56,744 (28,372  inventors$\times$2 time periods) in all the regressions. The inventors included have up to at least the 5th indirect collaborators as well as have information on the belonging establishments and firms.
(ii) The dependent variable is log of the average pairwise productivity of an inventor, $\ln y_{it}$, defined in terms of forward citations  in Panel A and novelty of patents in Panel B.
(iii) Explanatory variables shown are the average differentiated knowledge of collaborators, $\ln k_{it}^D$; the number of inventors in the firm and establishment of an inventor, $\ln f_{it}$ and $\ln e_{it}$, respectively; the scope of developing patents at the firm and establishment levels in terms of IPC subgroups, $\ln s_{it}^f$ and $\ln s_{it}^e$, respectively; the first- and second-order effects of the cumuraltive research scope, $\ln k_{it}$, of an inventor.
(iv) In all the regressions, $\ln k_{it}^D$ is treated as an endogenous variable, and is instrumented by the same variable of the 3rd-5th indirect collaborators of inventor $i$.
(v) In all the regressions, year and IPC class fixed effects, as well as a variety of neighborhood effects are controlled. Neighborhood effects include the sizes of concentration of inventors, R\&D expenditure, manufacturing employment and output within a circle of 1-km radius, and population within a circle of 20-km radius around a given inventor. 
(vi) The fourth row in each panel reports \cite{Olea-Pflueger-JBES2013}'s effective first-stage $F$-statistic.
(vii) The fifth row in each panel indicates 5\% critical value of the effective $F$-statistic with a Nagar bias threshold $\tau=10\%$. 
(viii) The last row in each panel indicates the $p$-value of \cite{Hansen-ECTA1982}'s $J$ test.
(ix) Robust standard errors in parentheses are clustered on urban agglomeration.
\end{tablenotes}
\end{table}

Tables \ref{tb:first-firm} and \ref{tb:firm-factors} summarize the first- and second-stage 2SLS-IV regression results for the key variables.
Panels A and B in the table show the IV results for \eqref{eq:baseline} under quality- and novelty-adjusted measures, respectively, with these additional controls on the RHS. The IVs are constructed using all the 3rd--5th indirect collaborators because similar results are obtained when only one of them is used.%
\footnote{The effective first-stage $F$ values, $F^{\rm{eff}}$, for the weak IV test by \cite{Olea-Pflueger-JBES2013} are reasonably large for all cases, indicating the strong relevance of the IVs. \cite{Hansen-ECTA1982}'s $J$-test indicates no evidence against the exogeneity of the IVs.}\ 

We find that the estimated coefficients of $\ln k_{it}^D$ in \eqref{eq:baseline} are different between the baseline and the current specifications (under both the quality- and novelty-adjusted productivity) when the size and research scope at the firm/establishment level are controlled.
But, the differences in all the specifications considered in Table \ref{tb:firm-factors} are within 10\% of the estimated coefficient in the baseline specification.
Thus, the effect of knowledge exchange on the collaborative productivity appears to be at most mildly influenced by the scale of a firm and an establishment.
The size effect is negative after controlling for the scope effect for both the firm and establishment levels and under both quality- and novelty-adjusted productivities.
It may reflect the fact that a larger firm/establishment tends to host a larger share of less experienced inventors. 
\smallskip

\noindent {\it Counterfactual collaboration} -- 
Next, we consider possible influence from broader neighborhood in the research network beyond the own firm or establishment.
Even though we focus on inventors whose firm affiliation is fixed throughout the study period, the patents are often developed jointly by an affiliation of multiple firms and establishments.
Affiliation-specific factors, such as the amount of available research funds, the presence of star inventors and spillovers, may influence collaborative output of inventors.


To evaluate the influence of the research affiliation of firms,
we consider random counterfactual choices of collaborators for each inventor in $I$ conditional on the actual number of their collaborators as well as firm/establishment affiliation of each.
Suppose that among the $n_{it}$ collaborators of inventor $i\in I$ in period $t$, $n_{it}^A$ belongs to firm $A$, $n_{it}^B$ belongs to firm $B$, and so on, where $n_{it} = \sum_j n_{it}^j$.
Then, these $n_{it}$ collaborators are replaced by $n_{it}^A$ randomly chosen collaborators without replacement (according to the uniform probability distribution) from firm A, $n_{it}^B$ from firm B, and so on; however, the second or closer indirect and direct collaborators of $i$ are excluded from the selection to mitigate the reflection problem.
We construct 1,000 sets of counterfactual collaboration patterns in this way and compute the counterfactual value $\tilde{k}_{it}^D$ under each.
Model \eqref{eq:baseline} is then estimated by OLS with and without firm-level controls, $\ln f_{it}$ and $\ln s_{it}^f$, using the counterfactual value of $\tilde{k}_{it}^D$.
Alternatively, a similar exercise is done for counterfactual affiliations among establishments rather than firms.

\begin{figure}[h!]
\includegraphics[scale=.2]{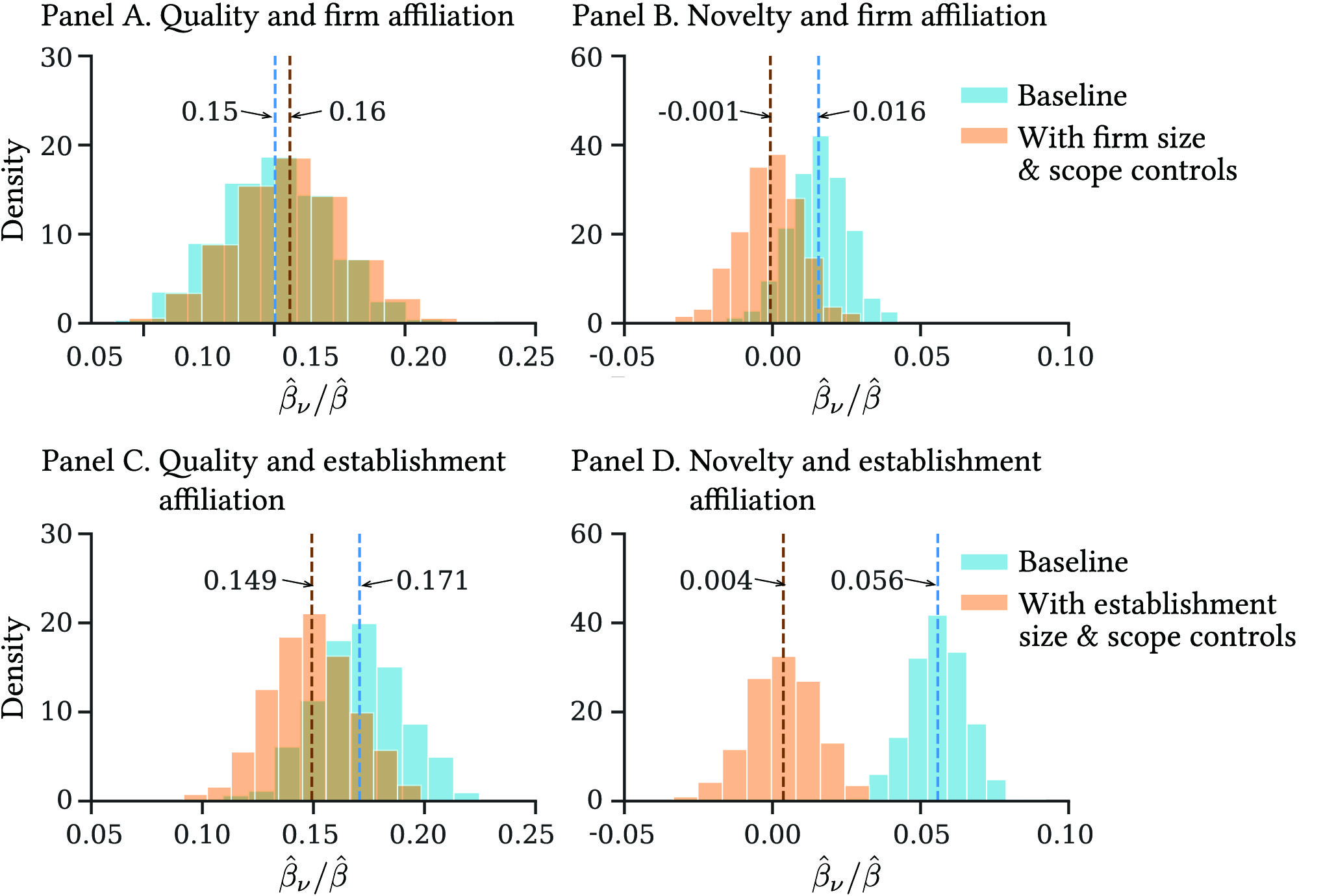}
\caption{Contribution of firm/establishment affiliations to collaborative productivity\label{fig:placebo}}
\begin{figurenotes}[\fontsize{10pt}{10pt}\selectfont Notes]
\fontsize{10pt}{10pt}\selectfont 
Each panel shows the distribution of the relative size, $\hat{\beta}_\nu/\hat{\beta}\:(\nu = 1,\ldots,1000)$, of the estimated coefficients of $\ln k_{it}^D$ between the random counterfactual and actual collaborations.
Inventors in each counterfactual collaboration are randomly selected conditional on the affiliation of inventors with firms or establishments in the actual collaboration.
The values of $\hat{\beta}_\nu$ are estimated by the OLS, whereas $\hat{\beta}$ is estimated by the IV regression using instruments constructed from all the 3rd-5th indirect collaborators based on model \eqref{eq:baseline} with and without controls for size and research scope of firms or establishments.
Panels A and B (C and D) show the cases under quality- and novelty-adjusted productivities, respectively, with and without controls for size and research scope at the firm (establishment) level.
\end{figurenotes}
\end{figure}

We interpret the estimated coefficient, $\tilde{\beta}$, for counterfactual $\ln \tilde{k}_{it}^D$ as 
the effect of common factors within an affiliation of firms/establishments.
It is then compared with the estimated coefficient, $\hat{\beta}$, under the actual collaborations based on IV regression using the instruments based on all the 3rd-5th indirect collaborators.
Figure \ref{fig:placebo} shows the distribution of the estimated shares, $\tilde{\beta}/\hat{\beta}$, for each pair of actual and counterfactual specifications of the model.

Regarding the firm-based counterfactual collaboration, as depicted in Panel A (Panel B) of Figure \ref{fig:placebo}, the share $\tilde{\beta}/\hat{\beta}$ takes values 0.15 and 0.16 (-0.001 and 0.016) on average under the productivity measured by quality (novelty) with and without controls for firm size and research scope, respectively.
In Panel C (Panel D) of Figure \ref{fig:placebo}, the corresponding values for the establishment-based counterfactual collaborators are found to be 0.15 and 0.17 (0.004 and 0.056), respectively.%

For the quality of collaborative output, firm and establishment-based counterfactual collaborations appear to account for a non-negligible portion of the positive effect of knowledge exchange.
Yet, on average more than 80\% of the effect remains to have direct impact on actual collaborative output.
For the novelty of collaborative output, they  appear to have only marginal effects.

\newpage
\section{\bf \thesection. Quantity-Quality and Quantity-Novelty Decomposition of Collaborative Productivity\label{app:qq}}
\noindent Tables \ref{tb:qq-decomposition-citation} and \ref{tb:qq-decomposition-novelty} show the second-stage result of the 2SLS-IV regressions for \eqref{eq:quality-equation} under quality- and novelty-adjusted productivities, respectively.
\begin{table}[h!]
\centering
\caption{Quality-quantity decomposition of collaborative productivity \label{tb:qq-decomposition-citation}}
\begin{tabular*}{\textwidth}{p{1in}@{\extracolsep{\fill}}ccccc} \midrule[0.6pt]\midrule[0.6pt]
& \multicolumn{4}{c@{}}{Panel A. Average quality} \\[2pt]
\cmidrule[0.6pt]{2-6}
Variable & (1) OLS & (2) IV3-5 & (3) IV3 & (4) IV4 & (5) IV5\\
\midrule[0.6pt]
$\ln k_{it}^D$ & 0.0823 & 0.0579 & 0.0577 & 0.0806 & 0.0759 \\
 & (0.00617) & (0.0331) & (0.0334) & (0.0522) & (0.107) \\
$\ln k_{it}$ & 0.0334 & 0.0368 & 0.0368 & 0.0337 & 0.0343 \\
 & (0.0180) & (0.0186) & (0.0186) & (0.0194) & (0.0233) \\
$(\ln k_{it})^2$ & -0.00276 & -0.00385 & -0.00386 & -0.00284 & -0.00304 \\
 & (0.00546) & (0.00566) & (0.00566) & (0.00595) & (0.00729) \\
\midrule[0.6pt]
$R^2$ & 0.054 & 0.054 & 0.054 & 0.054 & 0.054 \\
$F^{\mbox{eff}}$ &  & 534.7 & 1495 & 599.6 & 146.3 \\
\multicolumn{1}{l@{}}{Critical $F^{\mbox{eff}}$-value} &  & 9.281 & 23.11 & 23.11 & 23.11 \\ 
\multicolumn{1}{l@{}}{Hansen $J$ $p$-value} &  & 0.833 &  &  &  \\ 
\midrule[0.6pt]
\\
& \multicolumn{4}{c@{}}{Panel B. Quantity} \\[2pt]
\cmidrule[0.6pt]{2-6}
Variable & (1) OLS & (2) IV3-5 & (3) IV3 & (4) IV4 & (5) IV5\\
\midrule[0.6pt]
$\ln k_{it}^D$ & 0.0837 & 0.276 & 0.282 & 0.311 & 0.412 \\
 & (0.00504) & (0.0277) & (0.0281) & (0.0450) & (0.0972) \\
$\ln k_{it}$ & 0.129 & 0.103 & 0.102 & 0.0984 & 0.0847 \\
 & (0.0140) & (0.0149) & (0.0149) & (0.0158) & (0.0201) \\
$(\ln k_{it})^2$ & -0.0716 & -0.0631 & -0.0628 & -0.0615 & -0.0570 \\
 & (0.00431) & (0.00459) & (0.00461) & (0.00489) & (0.00639) \\
$R^2$ & 0.081 & 0.027 & 0.024 & 0.006 & -0.075 \\
$F^{\mbox{eff}}$ &  & 534.7 & 1495 & 599.6 & 146.3 \\
\multicolumn{1}{l@{}}{Critical $F^{\mbox{eff}}$-value} &  & 9.272 & 23.11 & 23.11 & 23.11 \\
\multicolumn{1}{l@{}}{Hansen $J$ $p$-value} &  & 0.300 &  &  &  \\
 \midrule[0.7pt]
\end{tabular*}
\begin{tablenotes}
\fontsize{10pt}{10pt}\selectfont
(i) The number of observations is 58,574 (29,287 inventors$\times$2 time periods) for all regressions. The inventors included have up to at least the 5th indirect collaborators in order to construct instruments for an endogenous variable. 
(ii) In Panel A, the dependent variable is log of the average patent quality of pairwise collaboration by inventor $i$, $\ln y_{it}^q$. In Panel B, it is log of the average number of patents developed in pairwise collaboration by inventor $i$.
(iii) The explanatory variables shown are the average differentiated knowledge of collaborators, $\ln k_{it}^D$; the first- and second-order effects of the research scope, $\ln k_{it}$, of an inventor.
(iv) Column 1 shows the result from the OLS regression, while columns 2-5 show the results of IV regressions, where $\ln k_{it}^D$ is treated as an endogenous variable, and is instrumented by the same variable of the 3rd-5th, 3rd, 4th and 5th indirect collaborators of inventor $i$ in columns 2, 3, 4 and 5, respectively. 
(v) In all the regressions, inventor, year and IPC class fixed effects, as well as a variety of neighborhood effects are controlled. Neighborhood effects include the sizes of concentration of inventors, R\&D expenditure, manufacturing employment and output within a 1-km radius, and population within a 20-km radius around a given inventor. 
(vi) The fifth row in each panel reports \cite{Olea-Pflueger-JBES2013}'s effective first-stage $F$-statistic, $F^{\rm{eff}}$.
(vii) The sixth row in each panel indicates 5\% critical value of the effective $F$-statistic with a Nagar bias threshold $\tau=10\%$. 
(viii) The last row in each panel indicates the $p$-value of \cite{Hansen-ECTA1982}'s $J$ test.
(ix) Robust standard errors in parentheses are clustered on urban agglomeration.
\end{tablenotes}
\end{table}

\begin{table}[h!]
\centering
\caption{Novelty-quantity decomposition of collaborative productivity\label{tb:qq-decomposition-novelty}}
\begin{tabular*}{\textwidth}{p{1in}@{\extracolsep{\fill}}ccccc} \midrule[0.6pt]\midrule[0.6pt]
& \multicolumn{4}{c@{}}{Panel A. Average novelty} \\[2pt]
\cmidrule[0.6pt]{2-6}
Variable & (1) OLS & (2) IV3-5 & (3) IV3 & (4) IV4 & (5) IV5\\
\midrule[0.6pt]
$\ln k_{it}^D$& 0.157 & 0.312 & 0.310 & 0.355 & 0.378 \\
 & (0.00610) & (0.0252) & (0.0252) & (0.0330) & (0.0449) \\
$\ln k_{it}$ & 0.105 & 0.0782 & 0.0785 & 0.0707 & 0.0666 \\
 & (0.0203) & (0.0210) & (0.0210) & (0.0215) & (0.0222) \\
$(\ln k_{it})^2$ & -0.111 & -0.0990 & -0.0992 & -0.0956 & -0.0937 \\
 & (0.00625) & (0.00657) & (0.00657) & (0.00686) & (0.00730) \\
\midrule[0.6pt]
$R^2$ & 0.132 & 0.108 & 0.109 & 0.094 & 0.084 \\
$F^{\mbox{eff}}$ &  & 841.1 & 2436 & 1359 & 635.1 \\
\multicolumn{1}{l@{}}{Critical $F^{\mbox{eff}}$-value} &  & 9.297 & 23.11 & 23.11 & 23.1\\
\multicolumn{1}{l@{}}{Hansen $J$ $p$-value}  &  & 0.0934 &  &  &  \\
\midrule[0.6pt]
\\
& \multicolumn{4}{c@{}}{Panel B. Quantity} \\[2pt]
\cmidrule[0.6pt]{2-6}
Variable & (1) OLS & (2) IV3-5 & (3) IV3 & (4) IV4 & (5) IV5\\
\midrule[0.6pt]
$\ln k_{it}^D$ & 0.0604 & 0.168 & 0.168 & 0.157 & 0.116 \\
 & (0.00419) & (0.0178) & (0.0178) & (0.0243) & (0.0340) \\
$\ln k_{it}$ & 0.130 & 0.111 & 0.111 & 0.113 & 0.120 \\
 & (0.0140) & (0.0145) & (0.0145) & (0.0147) & (0.0152) \\
$(\ln k_{it})^2$ & -0.0705 & -0.0620 & -0.0619 & -0.0629 & -0.0661 \\
 & (0.00432) & (0.00460) & (0.00460) & (0.00476) & (0.00513) \\
\midrule[0.6pt]
$R^2$ & 0.078 & 0.054 & 0.054 & 0.059 & 0.072 \\
$F^{\mbox{eff}}$ &  & 841.1 & 2436 & 1359 & 635.1 \\
\multicolumn{1}{l@{}}{Critical $F^{\mbox{eff}}$-value} &  & 9.293 & 23.11 & 23.11 & 23.11\\
\multicolumn{1}{l@{}}{Hansen $J$ $p$-value}  &  & 0.171 &  &  &  \\
 \midrule[0.7pt]
\end{tabular*}
\begin{tablenotes}
\fontsize{10pt}{10pt}\selectfont
(i) The number of observations is 58,574 (29,287 inventors$\times$2 time periods) for all regressions. The inventors included have up to at least the 5th indirect collaborators in order to construct instruments for an endogenous variable. 
(ii) In Panel A, the dependent variable is log of the average patent quality of pairwise collaboration by inventor $i$, $\ln y_{it}^q$. In Panel B, it is log of the average number of patents developed in pairwise collaboration by inventor $i$.
(iii) The explanatory variables shown are the average differentiated knowledge of collaborators, $\ln k_{it}^D$; the first- and second-order effects of the research scope, $\ln k_{it}$, of an inventor.
(iv) Column 1 shows the result from the OLS regression, while columns 2-5 show the results of IV regressions, where $\ln k_{it}^D$ is treated as an endogenous variable, and is instrumented by the same variable of the 3rd-5th, 3rd, 4th and 5th indirect collaborators of inventor $i$ in columns 2, 3, 4 and 5, respectively. 
(v) In all the regressions, inventor, year and IPC class fixed effects, as well as a variety of neighborhood effects are controlled. Neighborhood effects include the sizes of concentration of inventors, R\&D expenditure, manufacturing employment and output within a 1-km radius, and population within a 20-km radius around a given inventor. 
(vi) The fifth row in each panel reports \cite{Olea-Pflueger-JBES2013}'s effective first-stage $F$-statistic, $F^{\rm{eff}}$.
(vii) The sixth row in each panel indicates 5\% critical value of the effective $F$-statistic with a Nagar bias threshold $\tau=10\%$. 
(viii) The last row in each panel indicates the $p$-value of \cite{Hansen-ECTA1982}'s $J$ test.
(ix) Robust standard errors in parentheses are clustered on urban agglomeration.
\end{tablenotes}
\end{table}


\clearpage
\printbibliography[heading=subbibliography,title={{\bf\large\textrm{REFERENCES FOR APPENDICES}}}]

\end{refsection}
\end{document}